\documentclass[aps,pra,amsmath,amssymb,reprint,superscriptaddress,showkeys]{revtex4-2}

\usepackage{graphicx}
\usepackage{xcolor}
\usepackage{amsmath}
\usepackage{amssymb}
\usepackage{braket}
\usepackage{subfigure}
 
\newcommand{\blue}[1]{\textcolor{blue}{#1}}

\begin{document}

\title{Controlling Quantum Chaos: Time-Dependent Kicked Rotor}

\newcommand{\RegensburgUniversity}{Institut f\"ur Theoretische Physik, Universit\"at Regensburg, D-93040 Regensburg, Germany}
\newcommand{\Steve}{Department of Physics and Astronomy, Washington State University, Pullman, WA USA}

\author{Steven Tomsovic}
\email{tomsovic@wsu.edu}
\affiliation{\RegensburgUniversity}
\affiliation{\Steve}
\author{Juan Diego Urbina}
\affiliation{\RegensburgUniversity}
\author{Klaus Richter}
\affiliation{\RegensburgUniversity}

\begin{abstract}

One major objective of controlling classical chaotic dynamical systems is exploiting the system's extreme sensitivity to initial conditions in order to arrive at a predetermined target state.  In a recent letter [{\it Phys.~Rev.~Lett.}~{\bf 130}, 020201 (2023)], a generalization of this targeting method to quantum systems was demonstrated using successive unitary transformations that counter
the natural spreading of a quantum state.
In this paper further details are given and an important quite general extension is established.  In particular, an alternate approach to constructing the coherent control dynamics is given, which introduces a new time-dependent, locally stable control Hamiltonian that continues to use the chaotic heteroclinic orbits previously introduced, but without the need of countering quantum state spreading.
Implementing that extension for the quantum kicked rotor generates a much simpler approximate control technique than discussed in the letter, which is a little less accurate, but far more easily realizable in experiments.   The simpler method's error can still be made to vanish as $\hbar \rightarrow 0$.

\end{abstract}


\maketitle

\section{Introduction}

An important discipline within the larger field of {\bf \it controlling classical chaos}~\cite{Ott06} goes by the name of targeting~\cite{Ott90, Shinbrot90, Kostelich93, Bollt95, Schroer97}.  There the exponential instability of a chaotic dynamical system is taken advantage of in an optimal way to arrive at a particular classical target state.  Quite small perturbations lead to extraordinarily different evolutions and if chosen wisely enable an arrival at a target state relatively quickly even if the target state is quite distant from anywhere near the evolution of the unperturbed system.  In essence, the lack of predictability, i.e. production of dynamical entropy, is being converted into a resource for the targeting.  The existence of a direct targeting analogy in a chaotic quantum dynamical system is not at all apparent since for quantum dynamics the predictability, production of dynamical entropy, and even reversibility behave in fundamentally different ways~\cite{Kosloff81, Shepelyansky83, Tomsovic16}.  For example, the production of a quantum dynamical entropy vanishes for isolated, unitarily evolving systems regardless of the classical dynamics underlying nature.  Nevertheless, it was recently demonstrated that an analog of classical targeting could be extended to the quantum realm where it is dubbed {\it optimal coherent chaotic quantum targeting}~\cite{Tomsovic23}.  

Clearly, optimal coherent targeting belongs to the general classification of a quantum control problem.  A great deal of the early work on quantum control was motivated by a desire to control chemical reactions~\cite{Tannor85, Brumer86, Judson92, Warren93} where, due to rapid dispersal, various laser induced parametric excitation schemes were introduced, including optimal control techniques~\cite{Peirce88, Kosloff89}; a survey of the theory and applications can be found in~\cite{Dong10} and a more recent pedagogical overview of optimal control theory in~\cite{James21}.  The body of quantum control work is enormous, but for the most part not oriented toward dealing explicitly with quantum chaos where the main idea would be to take maximal advantage of exponential sensitivity in the underlying classical chaotic dynamics along with the evolving local structure of the neighboring dynamics; see however the works~\cite{Gong05, Bitter17}. Nevertheless, a few of the works rely on concepts of interest here.  In~\cite{Sugawara03}, Sugawara describes `wave packet shaping', which is applied to the integrable Morse oscillator potential.  It gives a complicated time-dependent laser field that shifts a wave packet from its ground state location to another desired location. Another method makes use of phase space structural implications of Kolmogorov-Arnol'd-Moser theory~\cite{Kolmogorov54, Arnold63, Moser62}; see also Ref.~\cite{Madronero06}, which describes the creation of a non-dispersive electronic wave packet following a trajectory about which the local phase space dynamics has an approximately harmonic nature.  

There have been some forays into controlling specifically quantum chaotic systems, such as the quantization of turnstiles leading to chaotic wave packet revivals~\cite{Tomsovic97}, using weak control fields to quantize Ulam's control conjecture~\cite{Gruebele07}, and the creation of NOON states via chaos-assisted tunneling~\cite{Vanhaele21, Vanhaele22}.  None of these address general schemes for optimal coherent targeting in quantum chaotic systems, which if fully developed might offer some important capabilities.  Classical chaos, which is inherently ergodic, visits all possible states of a system.  It's control engenders the possibility of placing a system in desired, but otherwise quite difficult-to-access states.  In a quantum mechanical context, optimal coherent targeting, as presented in~\cite{Tomsovic23} and here, concerns mostly close-to-minimum-uncertainty states, however within that realm there is the potential for accessing exotic states, for example, with delicate desired phase relationships.  Furthermore, the use of particular orbits, tailored to accessing the desired system states, can have the property of arriving in as quick a time as dynamically possible, which minimizes control errors that build up with increasing time and which may be critical if quantum phase coherence can only be maintained over some relatively short time scale.  The general extension discussed here, i.e.~alternate method, fits squarely into the important province of quantum simulation~\cite{Bloch08}.

The approach of extending targeting to quantum systems in Ref.~\cite{Tomsovic23} relies on semiclassical reasoning and makes use of chaotic trajectories called heteroclinic trajectories~\cite{Poincare99} as a resource for constructing unitary propagation matrices that guide an initial quantum state to a final one in a strongly chaotic system.  The new element introduced in~\cite{Tomsovic23} that is not present in classical targeting is based on successive unitary transformations that counter the natural spreading of a quantum state under the influence of the dynamics.  In this paper a different approach is described for countering the quantum state spreading that is more general, easier to implement experimentally, and which lends itself to a variety of approximations.  It involves constructing a new time-dependent, locally stable control Hamiltonian. It is designed to possess the chaotic heteroclinic orbit of interest as a solution, but with a locally stable dynamics.  This obviates the need of countering quantum state spreading.  Ahead, it is applied again to the quantum kicked rotor.

This system has not only been of paradigmatic importance for the conceptual development of quantum chaos theory, but has also been experimentally realized on different platforms. The majority of these experiments address the complex quantum dynamics of cold atoms in a kicked optical lattice~\cite{Moore95, Ammann98, Oberthaler99, Ryu06, Chabe08, Manai15, Hainaut18, Sajjad22, Cao22}. Since the control protocol proposed in the present work essentially requires invoking a refined sequence of kicks, this class of experiments should allow for realizing optimal coherent targeting and serve as a testbed for this new version of quantum control. 

This paper is structured as follows: in Sec.~\ref{background} the basic ideas behind optimal coherent control are outlined and a brief description of the classical and quantum versions of the kicked rotor, a paradigm of chaos, is given.  The ensuing Sec.~\ref{alternate} introduces an alternative path to optimal coherent control that has some distinct advantages relative to the work in~\cite{Tomsovic23}. This is followed by an application of this procedure to the kicked rotor in Sec.~\ref{applqkr}, where the accuracy is discussed as well.  The final section summarizes the main conclusions of this work and gives a brief outlook, followed by appendices covering the details of constructing the building block for coherent control, $\widehat U_{M^{-1}}$, generally and control Hamiltonians for the kicked rotor specifically.

\section{Background}
\label{background}

\subsection{Optimal coherent quantum targeting}

\begin{figure}
\includegraphics[width=\columnwidth]{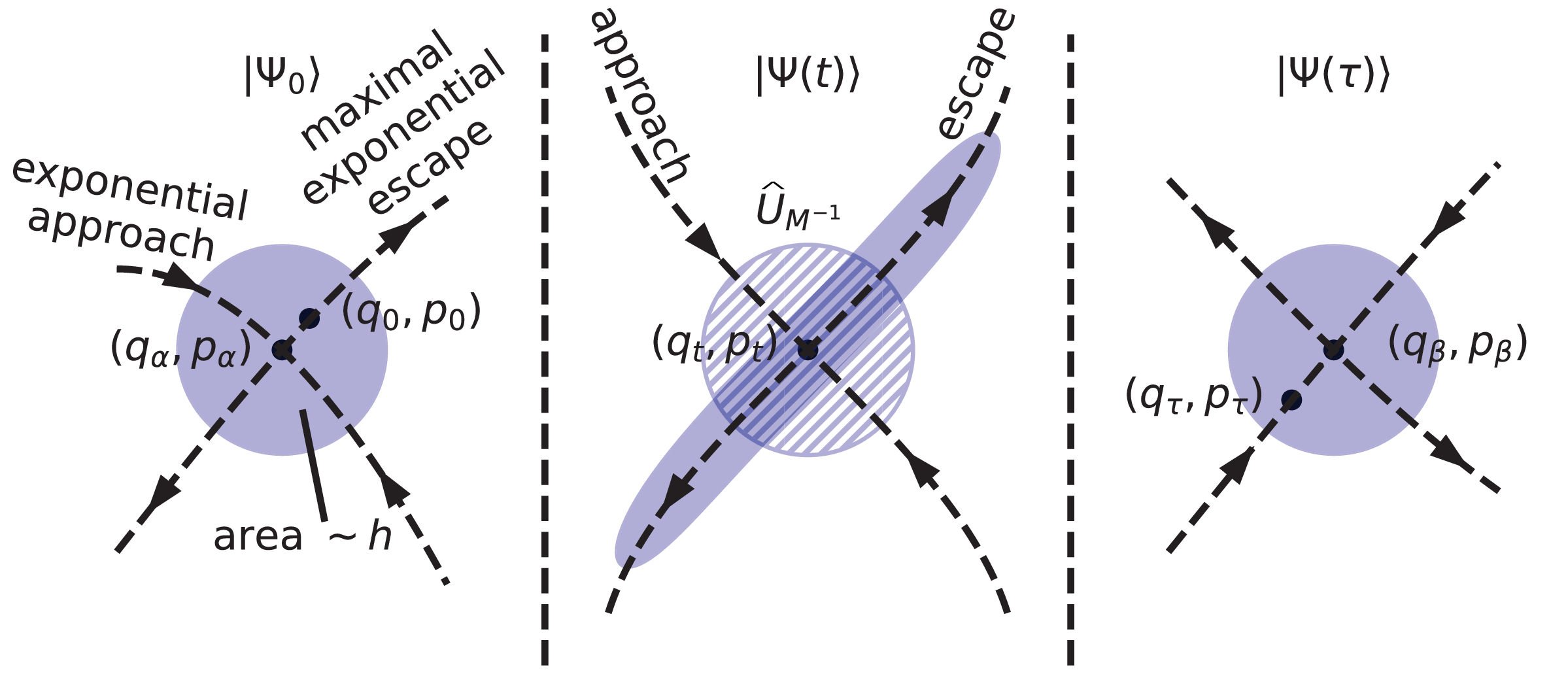}
\caption{Schematic of optimal coherent chaotic quantum targeting.  The circular zone represents the Wigner transform density of a minimal uncertainty state centered at $(q_\alpha,p_\alpha)$, which can be slightly shifted (via a unitary operator, $\widehat U_s$) to be centered on an optimal heteroclinic trajectory starting at $\left[{\bf p}_\gamma(0), {\bf q}_\gamma(0)\right]$ \big(designated $(q_0,p_0)$\big).  As it propagates the density follows this trajectory, but is locally spreading, which must be counteracted by  contractions, $\widehat U_{M^{-1}}$, see Eq.~(\ref{unt}). At the end, it can be shifted from $\left[{\bf p}_\gamma(\tau), {\bf q}_\gamma(\tau)\right]$ \big(designated $(q_\tau,p_\tau)$\big) to the centroid of the target state $(q_\beta,p_\beta)$
(reprinted with permission from 
Ref.~\cite{Tomsovic23})
. \label{fig1}}
\end{figure}

Starting with minimum uncertainty (or similarly localized) quantum states, i.e.~wave packets in ordinary quantum mechanics, Glauber coherent states~\cite{Glauber63} for bosons, it is possible to guide an initial state to some chosen final state following some chaotic (heteroclinic) trajectory of the quantum system's classical analog or mean field limit~\cite{Tomsovic23}.  Heteroclinic trajectories lie on the intersections of unstable and stable manifolds, have been used to construct semiclassical propagators in quantum chaotic systems~\cite{Tomsovic93, Tomsovic18}, and various methods exist  to locate them~\cite{Doedel89, Beyn90, Bai93, Liu97, Korostyshevskiy07, Li17}, even in several degree of freedom systems~\cite{Tomsovic18b}.  Ahead, the trajectory is labeled by a subscript $\gamma$.  Denoting the unitary dynamics of the uncontrolled quantum chaotic system of interest by $\widehat U(t)$, it is possible to create a controlled quantum dynamics given by
\begin{equation}
\label{unt}
\widehat U_{\rm CQD}(\tau) = \widehat U_s(\beta) \left[\prod_{n=0}^{l-1}\{\widehat U_{M^{-1}} \widehat U(t)\}_n\right]\widehat U_s(\alpha)\, ,
\end{equation}
where $\widehat U_{M^{-1}}$ unwinds the effects of localized quantum state's spreading, $\tau=lt$, and the $\widehat U_s$  are shift operators which displace the centroid of the initial state slightly towards the initial conditions of the heteroclinic orbit, $\left[{\bf p}_\gamma(0), {\bf q}_\gamma(0)\right]$, or the final trajectory point, $\left[{\bf p}_\gamma(\tau), {\bf q}_\gamma(\tau)\right]$,  towards the desired final state (target) centroid $\left(\widehat U_s(\Delta p,\Delta q)= \exp\left[i\left(\Delta p \hat q - \hat p\Delta q\right)/\hbar\right]\right)$.  In this way, an initially localized quantum state can be made to follow any trajectory segment that exists in the classical analog dynamical system; for a schematic illustration see Fig.~\ref{fig1} and for a proof of principle implementation see Ref.~\cite{Tomsovic23}. 

This technique of optimal coherent chaotic quantum targeting unwinds quantum state spreading from time to time by constructing a unitary operator, $\widehat U_{M^{-1}}$, associated with a linear canonical transformation.  The generators of $\widehat U_{M^{-1}}$ are at most quadratic in momentum and position operators or quadratures as the case may be~\cite{Moshinsky71}; see Appendix~\ref{appa} for details of its construction in a configuration space representation~\cite{Heller75,Heller91}.   In a sense, this is reminiscent of the use of Hessians in a broad class of large scale control problems~\cite{Nocedal06}.

Nevertheless, this may present the difficulty that although the $\widehat U_{M^{-1}}$ can be theoretically constructed, they may be rather difficult to physically realize.  For this reason, an alternative approach is discussed ahead which relies on a quite different, but designed Hamiltonian that does not require the construction of $\widehat U_{M^{-1}}$ and yet follows the desired chaotic heteroclinic trajectory of the original Hamiltonian.

\subsection{The usual quantum and classical kicked rotor}
\label{qckrb}

The kicked rotor, historically, has provided one of the most simple and powerful paradigms for both classical and quantum chaotic dynamical systems~\cite{Chirikov79, Fishman82, Shepelyansky83, Izrailev90, Lakshminarayan97, Tomsovic07}.  Its quantum version has been experimentally realized many times with cold atoms in a kicked optical lattice for a variety of purposes~\cite{Moore95, Ammann98, Oberthaler99, Ryu06, Chabe08, Manai15, Hainaut18, Sajjad22, Cao22}, and it has been realized in kicked molecular rotor systems~\cite{Bitter16} and analogous kicked light systems~\cite{Fischer99} as well. 

The usual kicked rotor Hamiltonian is given by
\begin{equation}
\label{krg}
H_0(p,q;t) = \frac{p^2}{2} - \frac{K}{4\pi^2}\cos (2\pi q) \sum_{n=-\infty}^\infty \delta(t-n) 
\end{equation}
which gives a single step quantum dynamics (from just before a kick to just before the next) generated by a unitary or Floquet operator,
\begin{equation}
\widehat{U}(1) = \exp\left( \displaystyle\frac{-i \widehat{p}^2}{2\hbar} \right) \; \exp\left[ \displaystyle\frac{iK}{4\pi^2\hbar}  \cos 2\pi\widehat{q} \right] \;.
\label{eq.4a3}
\end{equation}
In this form, the time between kicks has been scaled to unity and is not visible in the part of the unitary operator associated with free propagation kinetic energy.

Quantized on a unit phase space torus in a configuration representation, $U_{jk} = \langle q_j | \widehat{U}(1) | q_k\rangle$, with null Bloch phases it becomes
\begin{equation}
\resizebox{.9\hsize}{!}{$U_{jk} =\! \displaystyle\frac{1}{\displaystyle\sqrt{iN}} \exp\!\left[ \displaystyle\frac{i\pi (j\!-\!k)^2}{N} \right] \exp\!\left[ \displaystyle\frac{iNK}{2\pi}\! \cos\displaystyle\frac{2\pi k}{N} \right]$ ,}
\label{eq.4a4}
\end{equation}
where $N$ is the Hilbert space dimension, $j,k = 1,...N$, and Planck's constant is $2\pi\hbar = 1/N$ (in this equation unity represents the area of the fundamental torus being quantized).  The semiclassical limit of $\hbar \rightarrow 0$ is equivalent to $N\!\rightarrow \!\infty$.  

Minimum uncertainty Gaussian wave packets, $\ket{\alpha}$, can be expressed as
\begin{equation}
\label{gauss}
\braket{q_j | \alpha} = A(\hbar)\exp\!\left[\!- \frac{\left(q_j-q_\alpha\right)^2}{2\hbar} \!+\!\frac{i}{\hbar}p_\alpha \left(q_j\!-\!q_\alpha\right) \right]
\end{equation}
where $A(\hbar)$ is a normalization constant. This form shares momentum and position minimum uncertainty equally on the unit torus.  This wave packet's Wigner transform has a circular symmetry for all $\hbar$.  The area inside the wave packet's two standard deviation contour is equal to $h\ (=1/N)$, which is the area occupied by a single quantum state.  In this formula, the integer value $m$ of the position $q_j=m+j/N$ is chosen such that $\left| q_j-q_\alpha\right| \le 1/2$ for all $j$.  This is necessary because Eq.~\eqref{gauss} is an approximation and not represented in a periodic form.  It is valid as long as $\hbar$ is small enough that the two tails are quite small where they touch on the cylinder.  In this way, $p_\alpha = \braket{ \alpha| \widehat p |  \alpha}$  and likewise,  $q_\alpha = \braket{ \alpha| \widehat q |  \alpha}$.  For illustration ahead, the selected initial and target wave packet centroids are $(q_\alpha, p_\alpha)=(0.5,0)$ and $(q_\beta, p_\beta)=(0,0)$, respectively.

\begin{figure}
\includegraphics[width=\columnwidth]{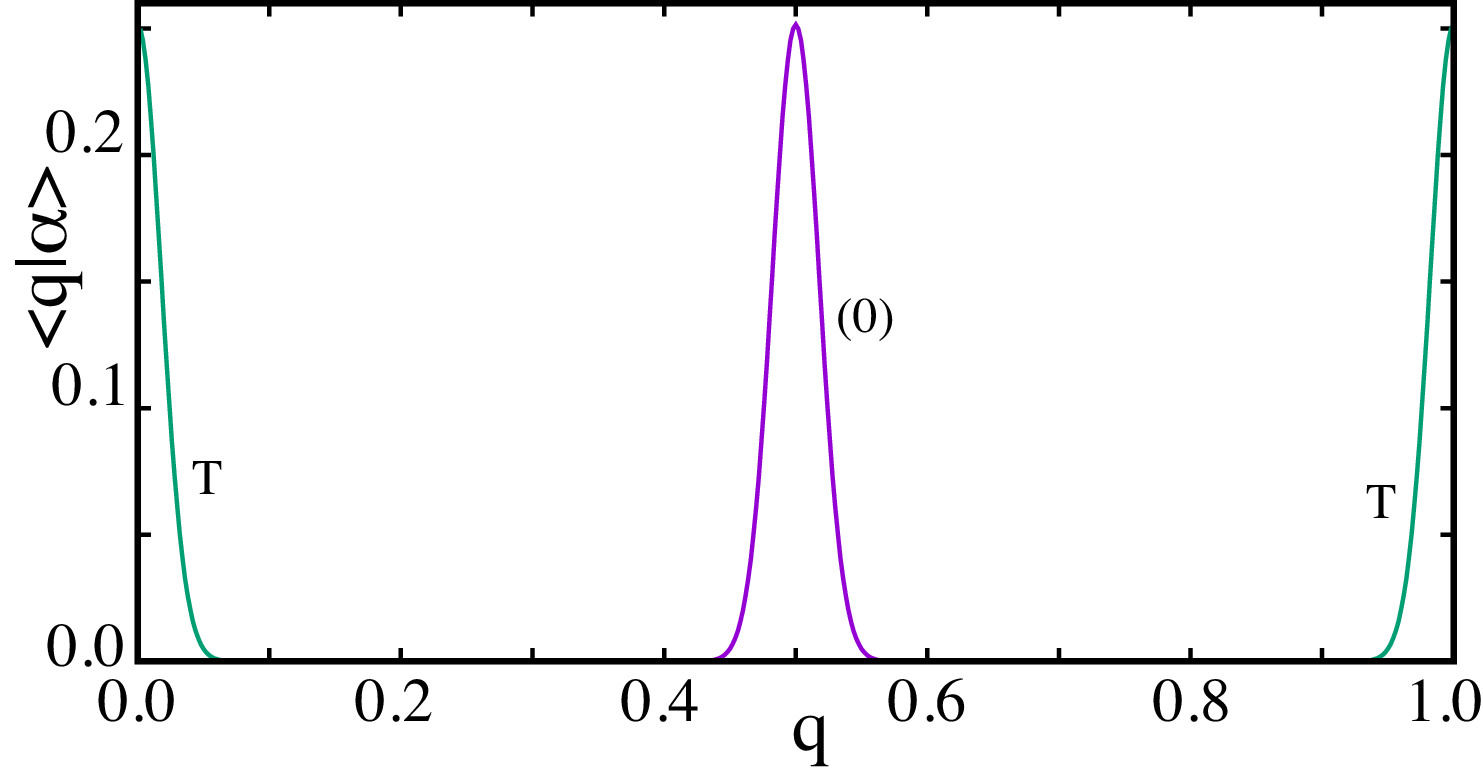}
\caption{Illustration of initial and target Gaussian wave packet states.  The dimensionality of the space is given by $N=200$ and thus $\hbar = 1/200\pi$.  The width is chosen so that the constant density contours of the Wigner transforms appear circular in Fig.~\ref{twa} ahead.  The initial state has momentum and position centroids of $(p_\alpha,q_\alpha)=(0.0,0.5)$ and the final state $(p_\beta,q_\beta)=(0.0,0.0)$.\label{fig2}}
\end{figure}

The resulting classical mapping equations are given by
\begin{equation} 
\label{eq:two}
\begin{split}
& p_{n+1} =p_{n}-\frac{K}{2\pi }\sin (2\pi q_{n}) \  \pmod 1 \ , \\
& q_{n+1} =q_{n}+p_{n+1}  \qquad\qquad \pmod 1  \ .
\end{split}
\end{equation} 
The associated single step stability matrix,
\begin{equation}
\left( \begin{array}{c} 
\delta p_{n+1} \\  
\delta q_{n+1}
\end{array}\right) = {\bf M}_n 
\left( \begin{array}{c} 
\delta p_{n}  \\ 
\delta q_{n} 
\end{array} \right) \; ,
\label{deltas2}
\end{equation}
is given by the product of matrices associated with the kick, $M_K$, and free propagation, $M_f$, contributions of the dynamics, respectively, i.e.
\begin{equation}
{\bf M}_f = \left( \begin{array}{cc} 
1 & 0 \\ 
1 & 1 \end{array} \right)\, , \, {\bf M}_K = 
\left( \begin{array}{cc} 1 & -K \cos \left(2 \pi q_n \right) \\
 0 & 1 \end{array} \right) \ ,
\label{deltas0}
\end{equation}
which after multiplication generates the form for a single time step
\begin{equation}
{\bf M}_n = \left( \begin{array}{cc} 
m_{11} & m_{12} \\ 
m_{21} & m_{22} \end{array} \right) = 
\left( \begin{array}{cc} 1 & -K \cos \left(2 \pi q_n \right) \\
 1 & 1-K \cos \left(2 \pi q_n \right) \end{array} \right) \, .
\label{deltas1}
\end{equation}
For large $K$-values, the dynamics are strongly chaotic, i.e.~the Lyapunov exponent is approximately $\mu=\ln\left(K/2\right)$~\cite{Chirikov79,Tomsovic07}.  This is all the classical information needed for the quantum targeting.  From the mapping, any trajectory can be calculated, in particular, the optimal heteroclinic one for the particular initial and final states.  For that trajectory, the stability matrix contains the information needed to construct the unitary transformations in Eq.~(\ref{unt}) that counteract the spreading.

\subsection{A small modification and example}
\label{small}

Ahead the controlled kicked rotor includes weak kicks at half time step intervals.  This alters a few details just discussed in Sec.~\ref{qckrb}.  In particular, the natural torus for quantization is no longer the unit square, but rather it encompasses double the momentum range.  The mapping equations for momentum utilize a $mod(2)$ instead of $mod(1)$.  Thus, the relation between the spatial dimension $N$ and $\hbar$ changes.  It becomes $N = S/2\pi\hbar$ where $S$ is the action (area) of the fundamental torus, which has doubled in value.  The relationship between spatial dimensionality and $\hbar$ becomes $N=1/\pi\hbar$ (i.e. half the states are associated with each unit area phase space cell).

To chose a concrete illustration for this paper, the initial and target states pictured in Fig.~\ref{fig2} provide  an excellent starting point and are identical to that of Ref.~\cite{Tomsovic23} (except for the change in the quantized torus).  Note that although these states happen to be centered on unstable fixed points of the mapping equations, this is only done for convenience and any two centroids could have been considered.

A new, longer optimal heteroclinic trajectory is chosen here for illustration whose coordinates are given in Table~\ref{orbit}.  This particular trajectory of all the six iteration heteroclinic ones has its initial condition and final endpoint closest to the centroids of the initial and final states, respectively.  The trajectory begins extremely close to the initial state  position and momentum centroids, and ends similarly close to the target state by a time, $\tau=6$, in units of the kicking period.    Hence, it is {\it the trajectory} of all possible trajectories whose shift operators, $\widehat U_s(\alpha),\ \widehat U_s(\beta)$, are closest to unit matrices.  In fact, the value of $\hbar$ would have to be very small for them to play an important role.  They are, nevertheless, fully accounted for ahead. 

The value of $K=8.00$ is selected for the rather strongly chaotic dynamics it generates with Lyapunov exponent $\mu=\ln 4$. Finally, a value of $\hbar=1/200\pi$ is used ahead because it is conveniently small enough to work well, but large enough for the example to exhibit some of the ways in which imperfections in the control Hamiltonian appear in the propagating state.
\begin{table}
\begin{center}
\begin{tabular}{|c|c|c|}
\hline 
 \quad time, $n$ \qquad  & \quad $q_n$ \quad  & \quad $p_n$ \qquad \\
\hline \hline 
 0   & 0.50060973724000 & 0.00054745314843 \\
 \hline 
 1   & 0.50603507637607 & 0.00542533913607 \\
\hline 
 2   & 0.55972945699483 & 0.05369438061876 \\
\hline 
 3   & 1.08012154306708 & 0.52039208607225 \\
\hline 
 4   & 0.98627390738657 & -0.09384763568051 \\
\hline 
 5   & 1.00209893780449 &0.01582503041792  \\
\hline 
 6   & 1.00113295252239 & -0.00096598528211 \\
\hline
\end{tabular}
\end{center}
\caption{The phase space coordinates of the chosen optimal trajectory for a time, $\tau=6$.  The modulus operation is not applied to the coordinates so that the fact that it wraps once around the torus in $q$ is visible.  Of the many $6$-iteration heteroclinic orbits, this particular one minimizes the two shift distances.
\label{orbit}}
\end{table}

\section{Alternate formulation of quantum targeting}
\label{alternate}

Instead of applying Eq.~\eqref{unt} and needing to construct and apply various $\widehat U_{M^{-1}}$, an alternative approach relies on the fact that given some particular solution of Hamilton's equations of motion, $\left[{\bf p}_\gamma(t), {\bf q}_\gamma(t)\right]$, for some particular initial conditions, $\left[{\bf p}_\gamma(0), {\bf q}_\gamma(0)\right]$, with the Hamiltonian, $H_0({\bf p},{\bf q};t)$, there exists an infinity of other Hamiltonians that possess that same particular solution.  Some of those other Hamiltonians have a stable dynamics in the neighborhood of the chosen solution.  For such a modified Hamiltonian, it is unnecessary to counteract quantum state spreading.  If one can identify a suitable new locally stable Hamiltonian, the $\widehat U_{M^{-1}}$ become irrelevant.

The first element of seeking a new Hamiltonian is to consider Hamilton's equations of motion for the particular trajectory.  The momentum and position variables can be replaced by their time dependent values, completely or partially, along the desired heteroclinic orbit.  In other words, any Hamiltonian which leads to the identical time sequence of values for $\left[\dot {\bf p}(t),\dot {\bf q}(t)\right]$, i.e.
\begin{eqnarray}
\label{controlH1}
\dot {\bf q}(t) &=& \frac{\partial H_0({\bf p}_\gamma(t), {\bf q}_\gamma(t); t)}{\partial {\bf p}_\gamma(t)} \, ,  \nonumber\\
\dot {\bf p}(t) &=& -\frac{\partial H_0({\bf p}_\gamma(t), {\bf q}_\gamma(t); t)}{\partial {\bf q}_\gamma(t)} \ ,
\end{eqnarray}
leads to the same solution for the initial conditions $\left[{\bf p}_\gamma(0), {\bf q}_\gamma(0)\right]$ and is a candidate for the control technique.  Amongst the infinite number of Hamiltonians which would respect Eq.~\eqref{controlH1} for the particular initial condition, the best choice would lead to a stable form for the stability matrix,  e.g.~the identity or a simple rotation matrix, and the control Hamiltonian being the simplest physically realizable one.  

A trivial example (presumably unphysical), not worrying about being the best or most optimal,  for a modified control Hamiltonian would be
\begin{align}
H_\gamma&({\bf p}, {\bf q}; t) =  \\
&\frac{\partial H_0({\bf p}_\gamma(t), {\bf q}_\gamma(t); t)}{\partial {\bf p}_\gamma(t)} \cdot {\bf p} + \frac{\partial H_0({\bf p}_\gamma(t), {\bf q}_\gamma(t); t)}{\partial {\bf q}_\gamma(t)}  \cdot {\bf q} \ ,\nonumber
\end{align}
but any other Hamiltonian leading to Eq.~\eqref{controlH1} is a suitable candidate.  Notice that $H_\gamma({\bf p}, {\bf q}; t)$ can be extraordinarily different from the original Hamiltonian, $H_0({\bf p}, {\bf q}; t)$, except in the neighborhood of the solution of interest $\left[{\bf p}_\gamma(t), {\bf q}_\gamma(t)\right]$.   Using any initial condition other than $\left[{\bf p}_\gamma(0), {\bf q}_\gamma(0)\right]$ would in all probability lead to completely different trajectories using the two different Hamiltonians, hence the $\gamma$ subscript on the Hamiltonian implies that it is designed exclusively to be used with the initial condition $\left[{\bf p}_\gamma(0), {\bf q}_\gamma(0)\right]$.

More generally for continuous dynamical systems, any time-dependent (elliptic) quadratic control Hamiltonian, $H_\gamma({\bf p}, {\bf q}; t)$, ensures that the stability matrix ${\bf M}_t\left[{\bf p}_\gamma(0), {\bf q}_\gamma(0)\right]$ retains a stable form.  This follows naturally from the stability matrix equation,
\begin{equation}
\label{stabmatH}
\frac{{{\rm d}\bf M}_t}{{\rm d}t} = \left(\begin{matrix}
-\frac{\partial^2 H({\bf p},{\bf q};t)}{\partial {\bf p} \partial {\bf q}} & -\frac{\partial^2 H({\bf p},{\bf q};t)}{\partial {\bf q} \partial {\bf q}} \\
\frac{\partial^2 H({\bf p},{\bf q};t)}{\partial {\bf p} \partial {\bf p}} & \frac{\partial^2 H({\bf p},{\bf q};t)}{\partial {\bf p} \partial {\bf q}} \\
\end{matrix}\right) {\bf M}_t
\end{equation}
Thus, an excellent starting point for identifying a desired $H_\gamma({\bf p}, {\bf q}; t)$ is to seek a time-dependent quadratic form subject to satisfying Eq.~\eqref{controlH1}.

Ahead in the application to the kicked rotor, the system can be considered to have a continuous dynamics interspersed with periodic kicks.  In such a case, the stability matrix follows by solving Eq.~\eqref{stabmatH} during the period $\tau$ between kicks, which  multiplies an initial kick stability matrix ${\bf M}_K$.  Hence, for one period from kick to kick, ${\bf M} = {\bf M}_\tau {\bf M}_K$.  It is this combined product that must remain stable over time for a suitable $H_\gamma({\bf p}, {\bf q}; t)$. The kicked rotor involves only free particle motion continously between kicks, which leads to a very simple form, but any dynamical system could be kicked and solved using Eq.~\eqref{stabmatH} and the product of stability matrices.

\section{Application to the quantum kicked rotor}
\label{applqkr}

\subsection{An efficient kicked rotor control Hamiltonian}
\label{krch}

It is indeed possible to identify a time-dependent, alternative kicked rotor which can be designed to follow some heteroclinic trajectory of the original kicked rotor, yet in a locally stable manner.  One method  adds a weak kick at half time steps in the process of identifying a modified, time-dependent kicked rotor that does not require the unwinding of quantum state spreading.  The sequence of logical steps leading to a modified Hamiltonian are detailed in Appendix~\ref{appb}.  Perhaps the simplest one is the example shown below, but others can be constructed, which in some cases lead to greater accuracy for the target state; again see Appendix~\ref{appb}.  In the limit of $\hbar\rightarrow 0$, this simple example is good enough to lead to a perfect arrival at the target state.  It is given by
\begin{equation}
\label{krgc}
\begin{split}
H_\gamma(p,q;t) = \frac{p^2}{2} + &\sum_{n=-\infty}^\infty  \bigg[ \delta(t-n) V_\gamma(q;n) + \\
&   \left. \delta\left(t-n-\frac{1}{2}\right) {\cal V}_\gamma\left(q;{n+\frac{1}{2}}\right) \right]
 \end{split}
\end{equation}
where
\begin{equation}
V_\gamma(q;n) = \frac{K}{4\pi^2}\sin\left[2\pi q_\gamma(n)\right]\sin\left(2\pi \left[q-q_\gamma(n)\right]\right)
\label{krv0}
\end{equation}
and
\begin{equation}
\label{krv1}
{\cal V}_\gamma\left(q;{n+\frac{1}{2}}\right) = -\frac{1}{5\pi^2}\cos\left(2\pi \left[q-q_\gamma\left(n+\frac{1}{2}\right)\right]\right)\, .
\end{equation}
A crude image of the intuitive idea behind the kick at the mid-point between strong kicks starts with the notion that during the free particle motion, the higher momentum components travel further than the lower momentum components.  If one can boost the lower momentum components, and diminish the higher momentum components just enough at the mid-time-point to offset this effect up to the midpoint, without altering the mean momentum, most of the spreading can be eliminated.  A more complete mathematical depiction of this basic idea is contained in the relations of Appendix~\ref{appb}.

\subsection{The classical controlled kicked rotor}
\label{cckr}

Instead of Eq.~\eqref{eq:two}, the modified Hamiltonian generates the mapping equations:
\begin{equation} 
\label{eq:map}
\begin{split}
& p_{n+\frac{1}{2}} =p_{n}-\frac{K}{2\pi }\sin\left[2\pi q_\gamma(n)\right]\cos\left(2\pi \left[q_n-q_\gamma(n)\right]\right) \\
&\qquad\qquad\qquad\qquad\qquad\qquad\qquad\qquad\qquad\qquad\ \pmod 2  \\
& q_{n+\frac{1}{2}} =q_{n}+\frac{p_{n+\frac{1}{2}}}{2} \qquad \qquad\qquad \qquad\qquad\qquad \pmod 1 \\
& p_{n+1} =p_{n+\frac{1}{2}}-\frac{2}{5\pi }\sin\left(2\pi\left[q_{n+\frac{1}{2}}-q_\gamma\left(n+\frac{1}{2}\right)\right]\right) \\
&\qquad\qquad\qquad\qquad\qquad\qquad\qquad\qquad\qquad\qquad\ \pmod 2  \\
& q_{n+1} =q_{n+\frac{1}{2}}+\frac{p_{n+1}}{2}  \qquad \qquad\qquad \qquad\qquad \quad\ \pmod 1  \, . \\
\end{split}
\end{equation}
For the initial conditions, $\left[{\bf p}_\gamma(0), {\bf q}_\gamma(0)\right]$, it is straightforward to see that these equations generate the same trajectory, $\left[{\bf p}_\gamma(t), {\bf q}_\gamma(t)\right]$ as Eq.~\eqref{eq:two} if a mod(2) is applied instead of the mod(1) for the momentum change.  Thus, the modified Hamiltonian successfully follows exactly the trajectory of interest.

On the other hand, the neighboring orbits are significantly modified, so much so that the stability matrix no longer reflects exponential instability locally.  To see this, note that the single step stability matrix can be constructed as the product of the two half step matrices.  Thus,
\begin{equation}
\label{m1m2}
{\bf M}_n(\gamma) = {\cal M}_{\frac{1}{2}}(2) \times {\cal M}_{\frac{1}{2}}(1)
\end{equation}
where
\begin{align}
\label{composite}
{\cal M}_{\frac{1}{2}}(2) &= \left( \begin{array}{cc} 
1 &  -\frac{4}{5}\cos\left(2\pi\left[q_{n+\frac{1}{2}}-q_\gamma\left(n+\frac{1}{2}\right)\right]\right)  \\ \\
\frac{1}{2}  & 1- \frac{2}{5}\cos\left(2\pi\left[q_{n+\frac{1}{2}}-q_\gamma\left(n+\frac{1}{2}\right)\right]\right)  \end{array} \right) \nonumber \\
{\cal M}_{\frac{1}{2}}(1) &= \left( \begin{array}{cc} 
1 & K\sin\left[2\pi  q_\gamma(n)\right]\sin\left[2\pi\left(q_n-q_\gamma(n)\right)\right]  \\ \\
\frac{1}{2}  & 1+ \frac{K}{2} \sin\left[2\pi  q_\gamma(n)\right]\sin\left[2\pi\left(q_n-q_\gamma(n)\right)\right] \end{array} \right)
\end{align}
For the initial condition, $\left[{\bf p}_\gamma(0), {\bf q}_\gamma(0)\right]$, the difference arguments of the cosine and sine functions vanish and the stability matrix reduces to
\begin{equation}
{\bf M}_n(\gamma) = \left( \begin{array}{cc} 
\frac{3}{5} &  -\frac{4}{5}  \\ \\
\frac{4}{5}  & \frac{3}{5}  \end{array} \right)
\label{modifiedM}
\end{equation}
which is a simple rotation matrix of fixed angle (roughly 53$^\circ$), independent of $n$.

\begin{figure}
\includegraphics[width=0.95\columnwidth]{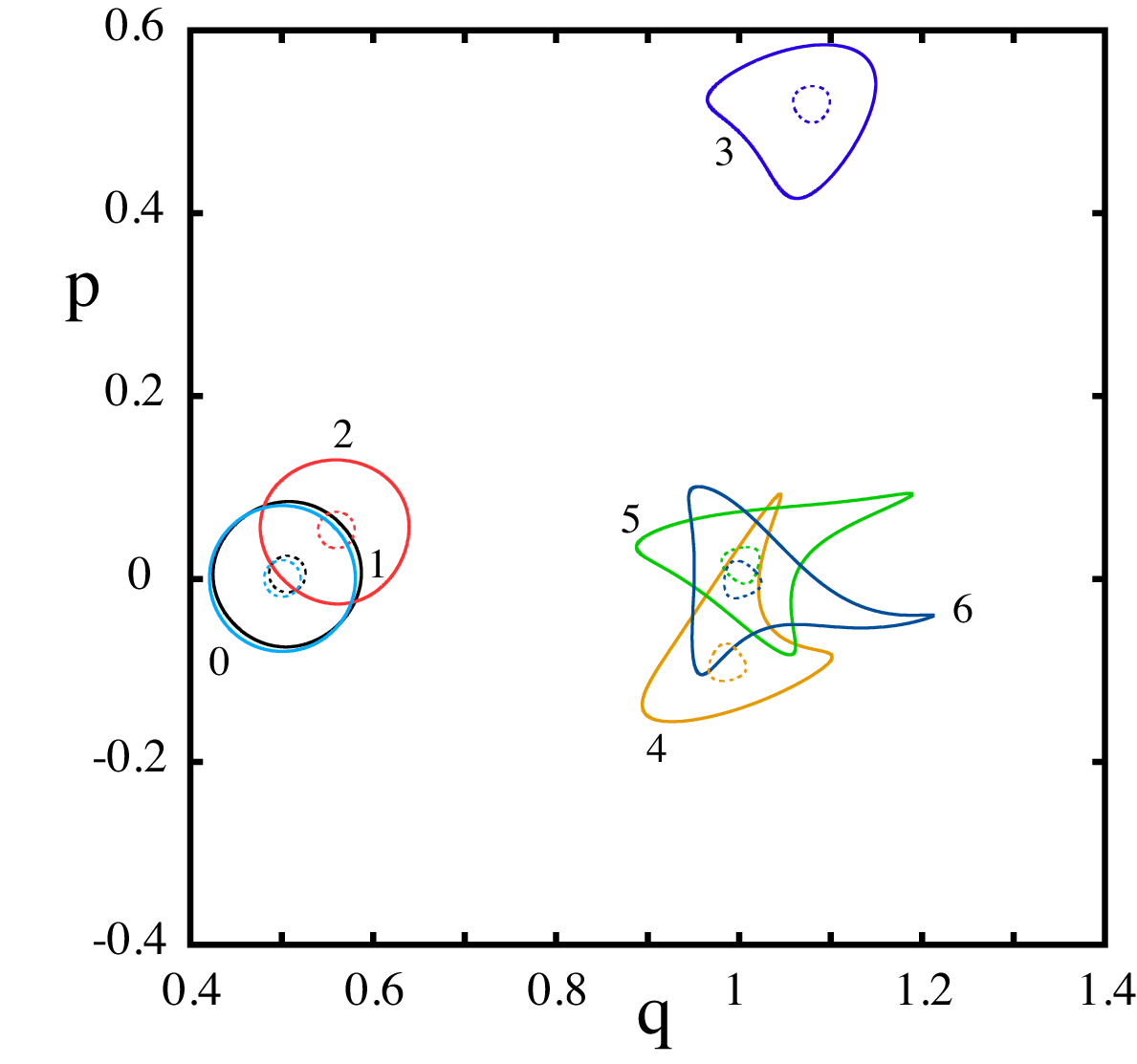}
\caption{Evolution of the Wigner transform densities of the propagated initial state.  The numbers label the time for each propagated density.  The outer circle is the $2\sigma$ contour, which encloses an area $h$, i.e.~that area occupied by a single quantum state.  The $\sigma/2$ contour is also drawn (inner dashed circles). \label{twa}}
\end{figure}

In the very close neighborhood of $\left[{\bf p}_\gamma(t), {\bf q}_\gamma(t)\right]$, there are trajectories for which the equalities
\begin{align}
0 &= \sin\left[2\pi\left(q_n-q_\gamma(n)\right)\right]\\
1 &= \cos\left(2\pi\left[q_{n+\frac{1}{2}}-q_\gamma\left(n+\frac{1}{2}\right)\right]\right) 
\end{align}
are no longer satisfied, but the deviations are nevertheless quite small, at least for some amount of time.  Such trajectories also remain stable in the neighborhood of $\left[{\bf p}_\gamma(t), {\bf q}_\gamma(t)\right]$.  However, the linearizable dynamical neighborhood is effectively smaller than in the case relying on $\widehat U_{M^{-1}}$ and the accuracy for a given value of $\hbar$ using the modified Hamiltonian is seen ahead to be not quite as good as in~\cite{Tomsovic23}.

The evolution of the controlled Wigner densities is illustrated in Fig.~\ref{twa}.  A number of features can be seen clearly.  First, note the exponential acceleration of the density's movement away from its starting point as time increases.  It is best seen by comparing the shifts in the densities between the $n=0,1,2, 3$ densities.  Conversely, there is an exponential compression seen by looking at the shifts between the $3,4,5,6$ densities.  Another feature to note is the increasing distortion of the densities $2\sigma$ contour as a function of increasing time.    In fact, the $\hbar$-value for the illustration is chosen to be small enough that the density is following the heteroclinic orbit properly, but large enough that the imperfect cancelling of the wave packet spreading is visible; if it were perfect each simple closed curve would remain a circle.  In addition, the improved quality of the method with shrinking $\hbar$ can be deduced from this figure as well.  The $\sigma/2$ contour, which is far less distorted from a perfect circle than the $2\sigma$ contour, can also be considered the $2\sigma$ contour for a value of $\hbar$ sixteen times smaller.  Thus, as $\hbar\rightarrow 0$ the modified kicked rotor discussed here does a better and better job of maintaining the local neighborhood (as measured by $\hbar$) as dynamically just a rotation.

\subsection{The quantum controlled kicked rotor}
\label{qckr}

The quantum unitary operator for the controlled kicked rotor, Eqs.~(\ref{krgc}-\ref{krv1}), following the desired heteroclinic trajectory of Table~\ref{orbit}, can be constructed in a straightforward way.  The matrix elements of the first and second half time steps in a configuration basis, $U_{jk}(1;n)$ and $U_{jk}(2;n)$, respectively, are given by
(with $q_j=j/N$)
\begin{widetext}
\begin{align}
U_{jk}(1;n) &= \frac{1}{\sqrt{iN}} \exp\left[ \frac{i\pi (j-k)^2}{N} \right] \exp\left(-\frac{iNK}{4\pi} \sin\left[2\pi q_\gamma(n)\right]  \sin\left[2\pi \left(q_j-q_\gamma(n)\right)\right] \right) \nonumber \\
U_{jk}(2;n) &= \frac{1}{\sqrt{iN}} \exp\left[ \frac{i\pi (j-k)^2}{N} \right] \exp\left(-\frac{iN}{5\pi} \cos\left[2\pi \left(q_j-q_\gamma\left[n+\frac{1}{2}\right]\right)\right] \right) \, .
\label{f0}
\end{align}

\begin{figure}
\includegraphics[width=0.33\textwidth]{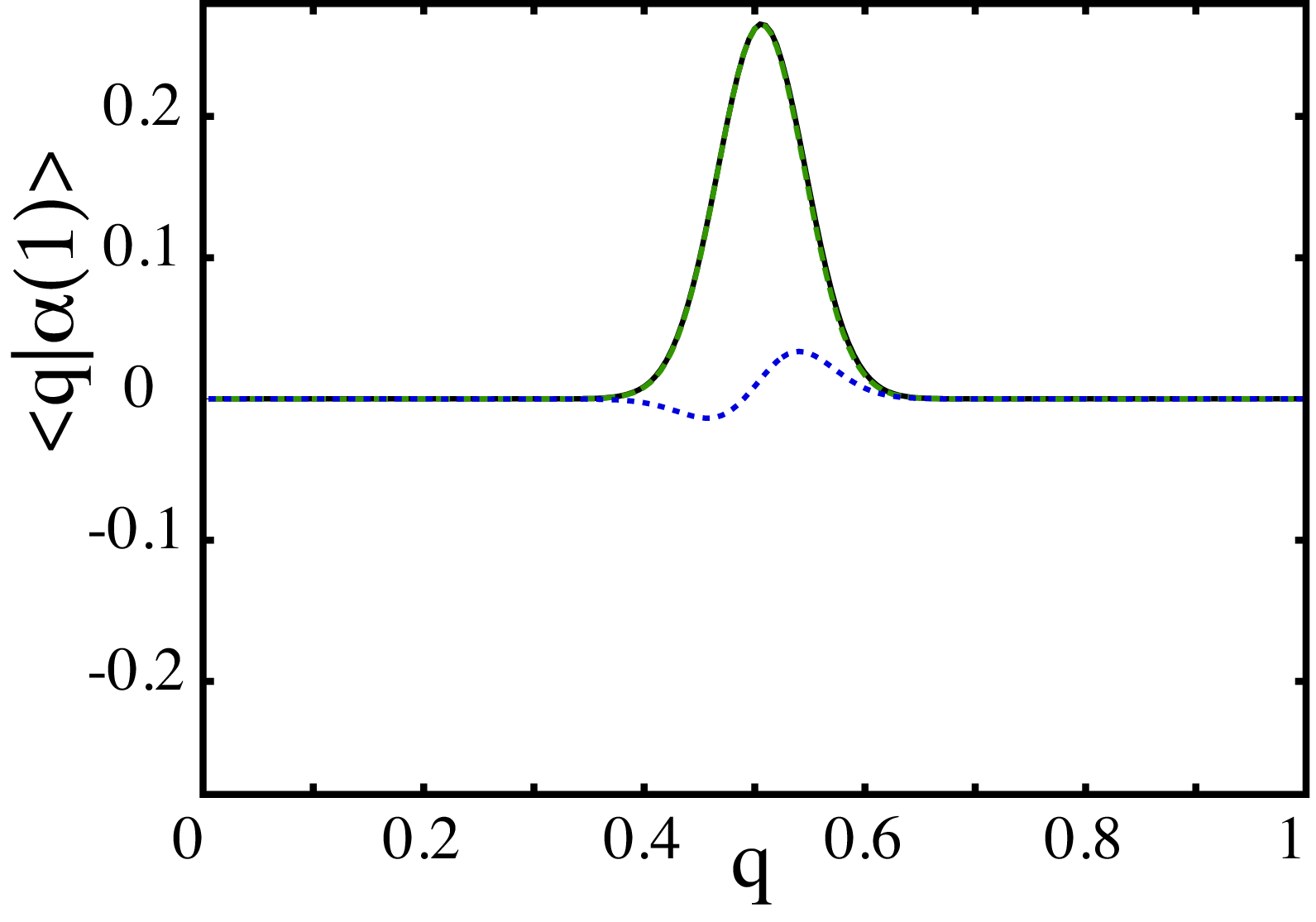}\,\includegraphics[width=0.33\textwidth]{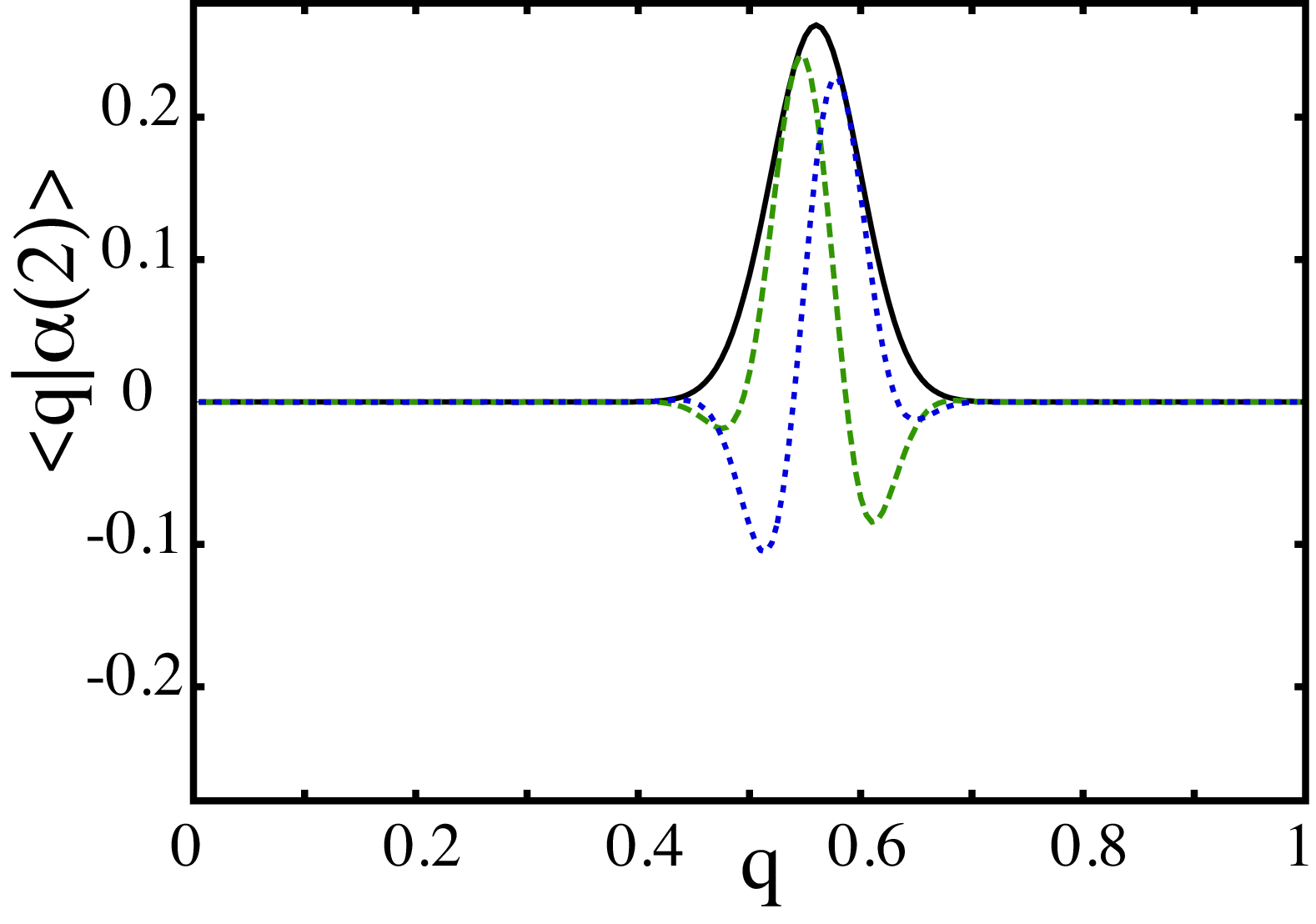}\,\includegraphics[width=0.33\textwidth]{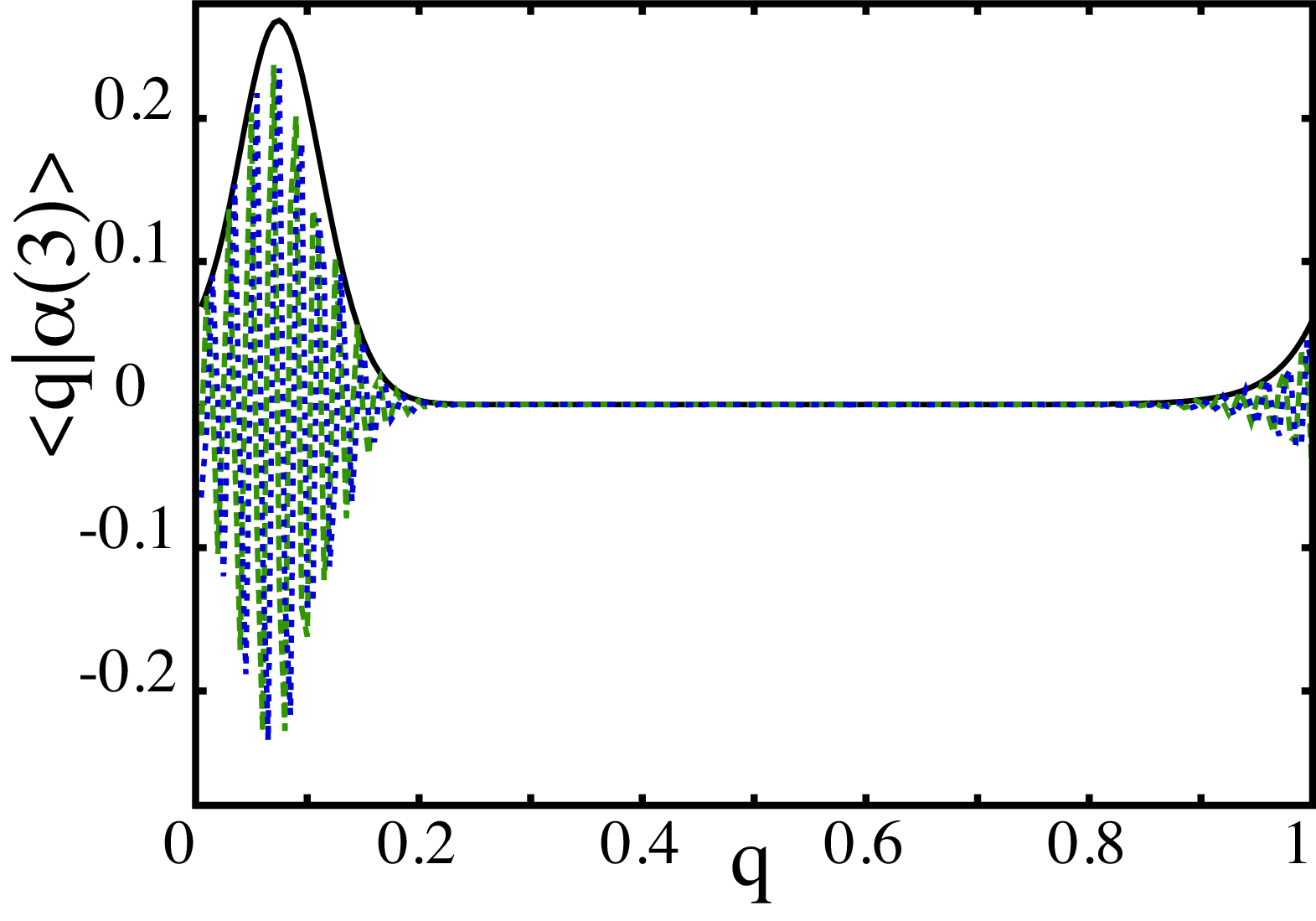}
\includegraphics[width=0.33\textwidth]{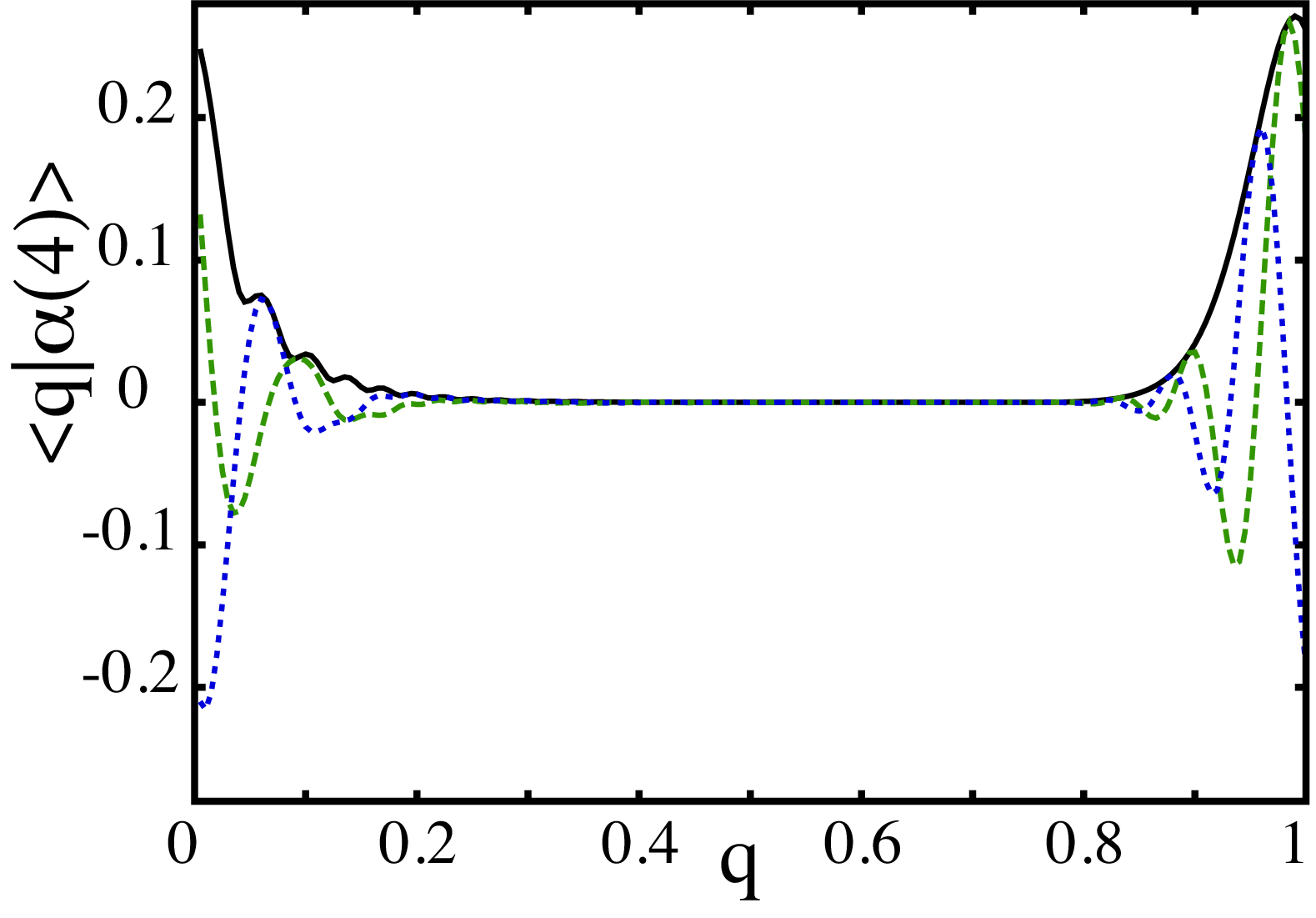}\,\includegraphics[width=0.33\textwidth]{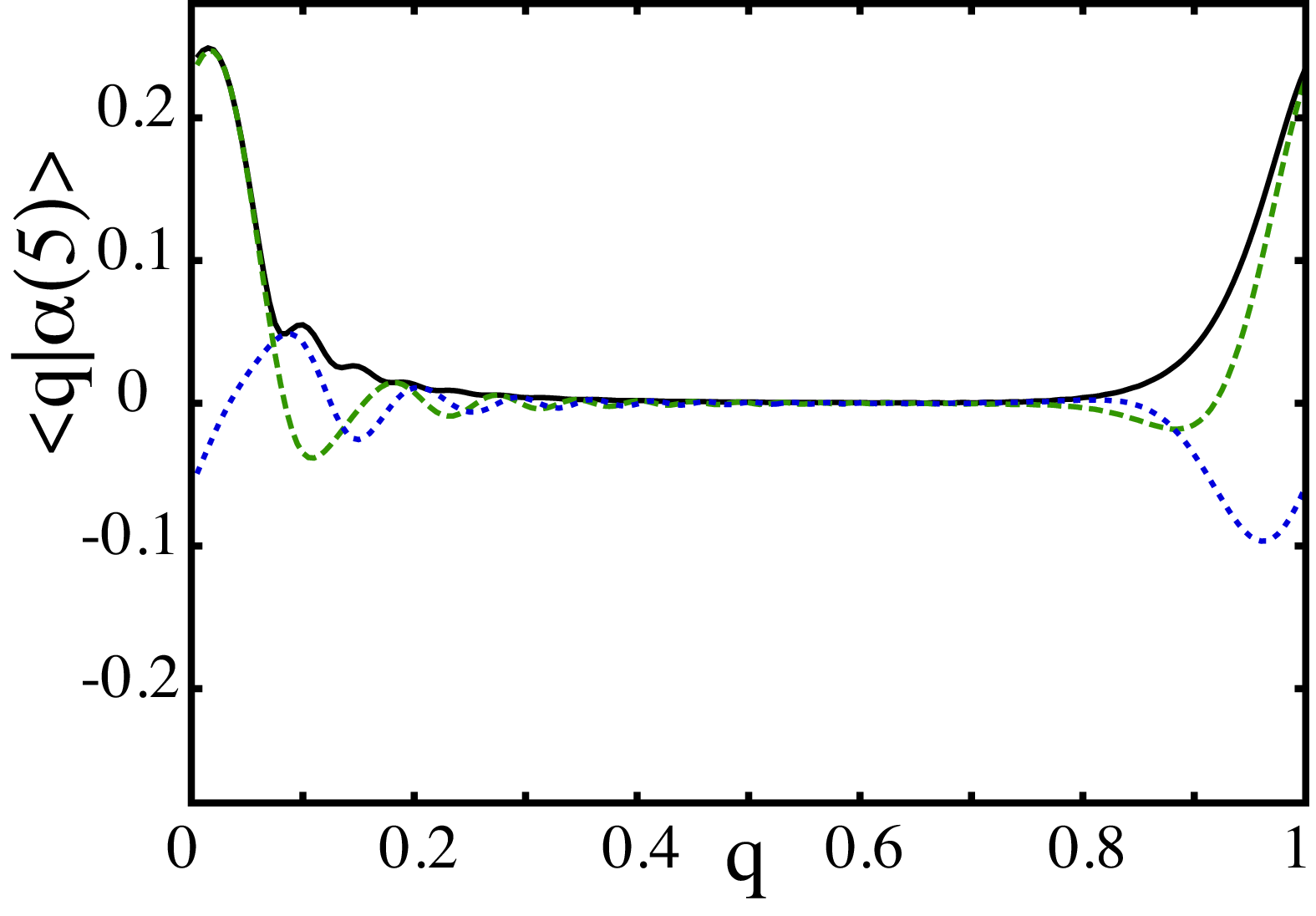}\,\includegraphics[width=0.33\textwidth]{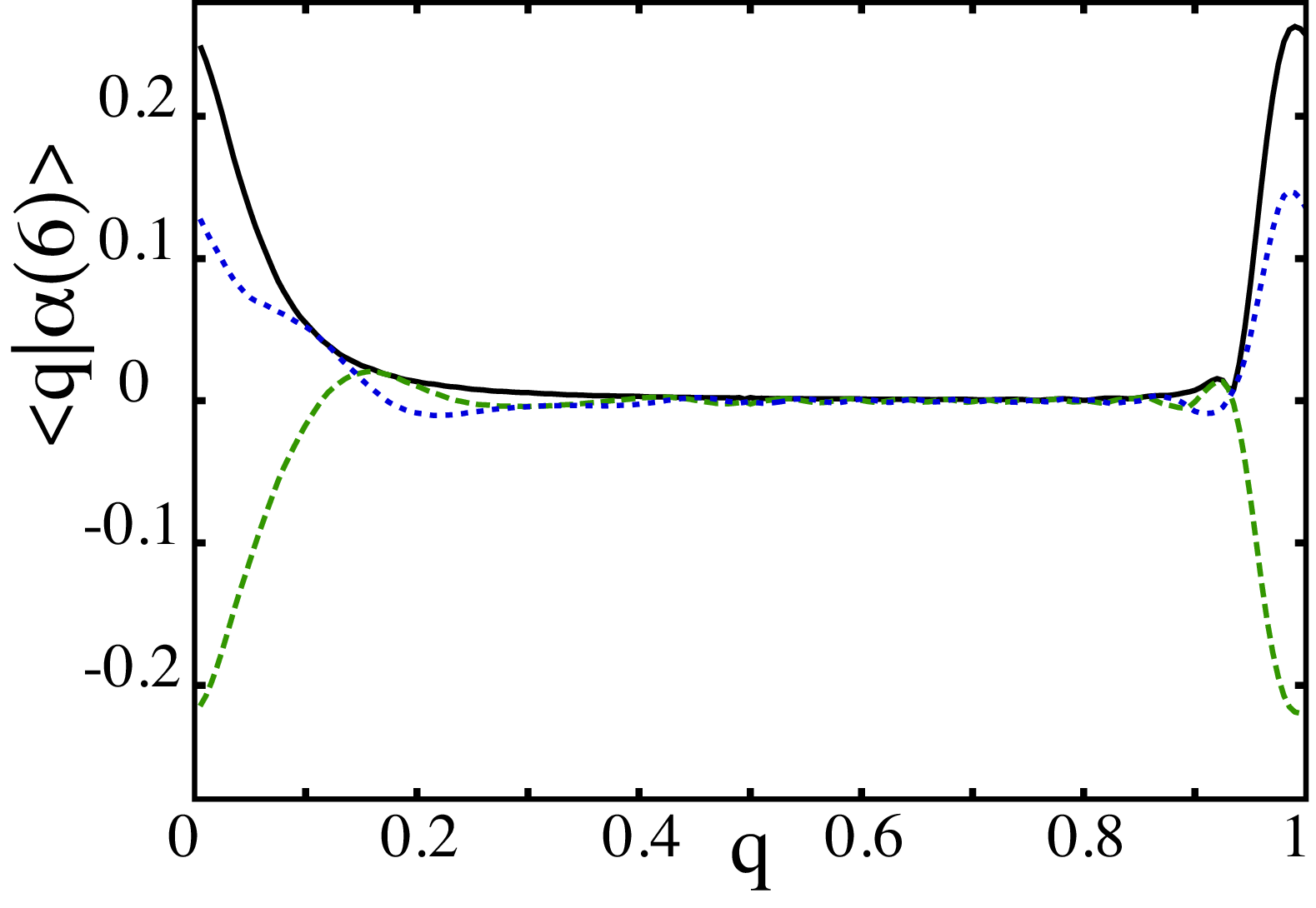}
\caption{Quantum propagation along the chosen heteroclinic trajectory  of Table \ref{orbit}.  The six panels from upper left to lower right mark the evolution at the trajectory's six full time steps.  The solid lines are the absolute value, the dashed (green) lines are the real part, and the dotted (blue) lines are the imaginary part of the controlled propagated wave function, $\braket{q|\alpha(t)}$.   As time increases, deviation from the ideal Gaussian wave packet becomes more visible.  Nevertheless, the state follows the heteroclinic trajectory in Table~\ref{orbit} quite closely.
\label{prop6}}
\end{figure}
\end{widetext}

The product, $\widehat U(2;n)\widehat U(1;n)$, generates a full time step unitary propagator, $\widehat  U_n(1)$.
The overall propagation is then given by
\begin{equation}
\widehat  U_\gamma(l) = \widehat  U_s(\beta)\left[ \prod_{n=0}^{l-1} \widehat  U_n(1) \right]\widehat  U_s(\alpha)
\end{equation}
where there is no longer a need for $\widehat  U_{M^{-1}}$, but the shift operators of Eq.~\eqref{unt} are kept.  Using this product of the constructed $\widehat  U_n(1)$ leads to the results shown in Fig.~\ref{prop6}.  The wave packet is clearly seen following the heteroclinic orbit, but beyond a time $t=3$, there appear some small magnitude oscillations in the tails that are undesirable and there are slower decaying tails than would be expected of a Gaussian wave packet.  These imperfections are traceable to the distortions of the Wigner density seen in Fig.~\ref{twa} for later times.

\begin{figure}[ht]
\begin{center}
\includegraphics[width=\columnwidth]{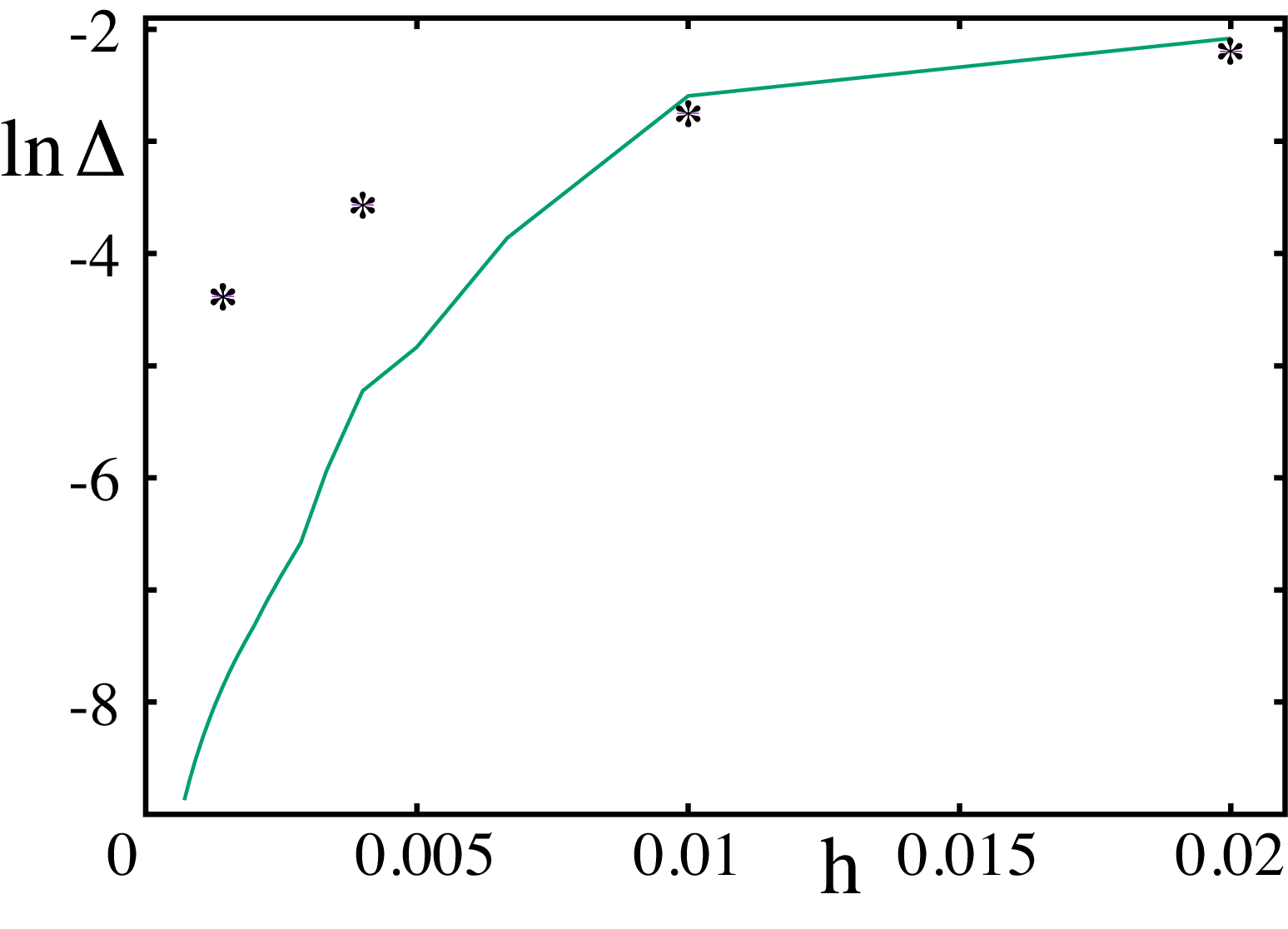}
\end{center}
\caption{Accuracy of quantum targeting for the kicked rotor.  The deviation of the squared overlap of the target state with the controlled propagated state from unity is plotted versus Planck's constant, $h$.  As $\hbar\rightarrow 0$, the error tends to vanish exponentially.  Dimensionalities ranging from $N=50$ to $1400$ are included.  The asterisks derive from the dynamics of the control Hamiltonian and various details of Sec.~\ref{applqkr}.  The line is drawn using the results of~\cite{Tomsovic23} and the $\widehat U_{M^{-1}}$ operators.
}
\label{overlap2}
\end{figure}

The accuracy of the final controlled targeting dynamics can explored by calculating the deviation from unity of the overlap $|\braket{\beta|\widehat U_\gamma(\tau=6)|\alpha}|^2$,
\begin{equation}
\Delta = 1- |\braket{\beta|\widehat U_\gamma(\tau)|\alpha}|^2 \, .
\end{equation}
The smaller $\Delta$, the greater the accuracy.  This deviation is plotted in Fig.~\ref{overlap2}, where the stars represent the results of Eqs.~(\ref{krgc}-\ref{krv1}) compared with the prior approach of~\cite{Tomsovic23} that more perfectly unwinds the wave packet spreading using the stability matrix to create the $\widehat U_{M^{-1}}$ operators.  Although, less perfect, this approach is presumably just as feasible as the original kicked rotor, as it requires only a weak extra kick halfway between the original kicks and control over the relative phase and strength of kicks.  It does not need the construction of $\widehat U_{M^{-1}}$ operators, which may be easy to construct theoretically, but possibly not in practice.  Finally, the errors disappear as $\hbar\rightarrow 0$, i.e. in this limit the accuracy still approaches perfection.

\begin{figure}[ht]
\begin{center}
\includegraphics[width=\columnwidth]{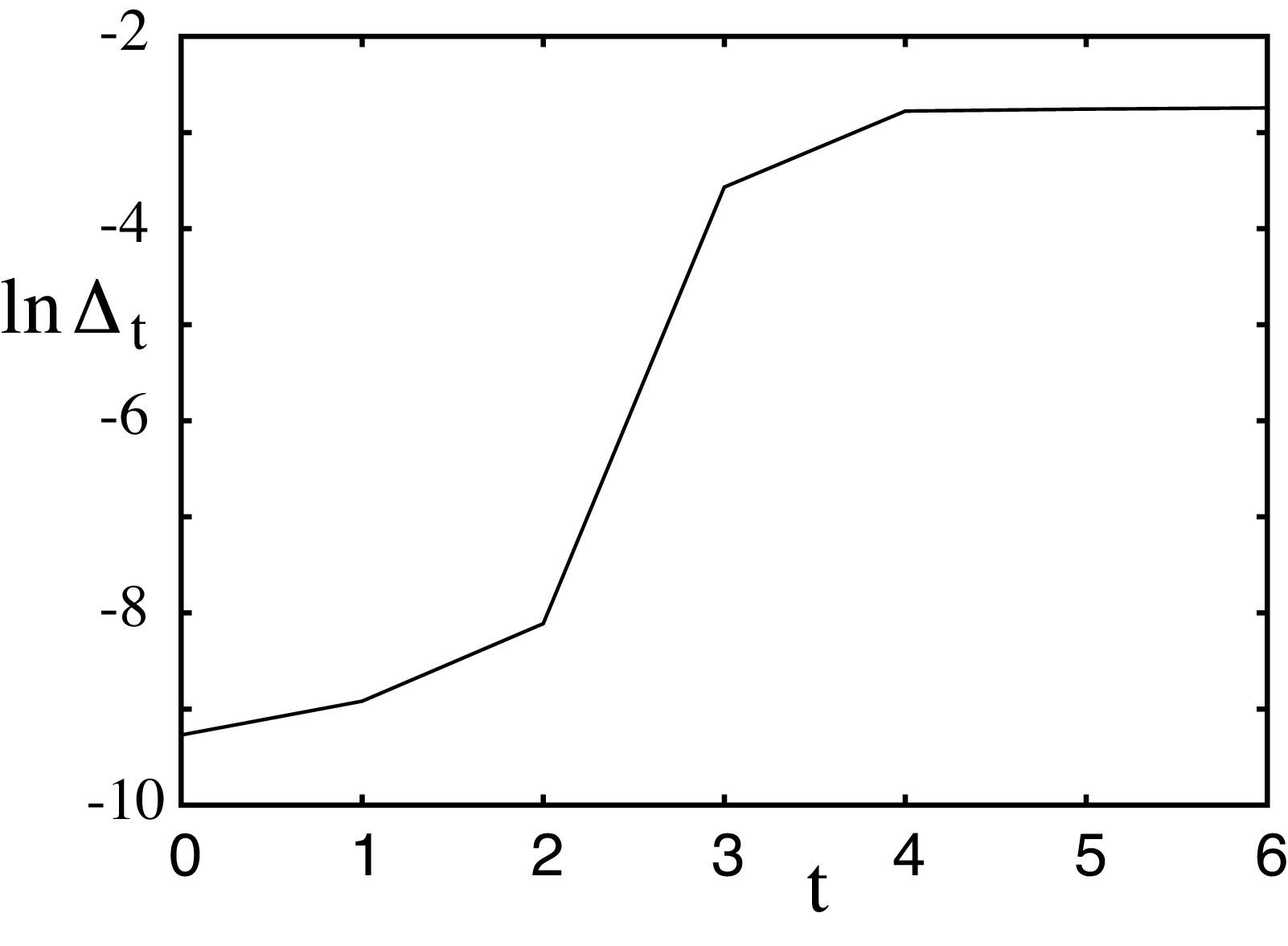}
\end{center}
\caption{Accuracy of the quantum targeting propagation for the kicked rotor for $h=0.01$ ($N=200$).  The deviation from unity of the squared overlap of a wave packet centered at $(q_t,p_t)$ ($t=n$, see Table~\ref{orbit}) with the controlled propagating state is plotted versus number of iterations.  Nearly all of the inaccuracy arises from the third and fourth iterations.  The final value at $t=6$ matches the point (star) at $h=0.01$ in Fig.~\ref{overlap2} as it must.
}
\label{fivehalf}
\end{figure}

It is also possible to follow the accuracy as a function of time.  If the controlled propagation were perfect, the propagated state would have unit overlap with a Gaussian wave packet, Eq.~\eqref{gauss}, centered at $(q_t,p_t)$.  Defining for this purpose, $\Delta_t$ as one minus the absolute square of this overlap at $t=n$ gives the result shown in Fig.~\ref{fivehalf}.  
There is negligible error in the initial shift, and the accuracy is nearly exclusively lost during the third and fourth iterations.  This is entirely consistent with Fig.~\ref{twa}.  Inspection there shows that nearly all of the distortion of the initial circular shape towards a roughly distorted triangular shape occurs from the same third and fourth iterations.  Thus, the classical Wigner transform propagation clearly holds the information about the accuracy with time in its propagating shape.

\section{Conclusions}

This paper gives both a generalization of the optimal coherent quantum targeting in classically chaotic systems presented in~\cite{Tomsovic23} and provides some additional details; e.g.~see Appendix~\ref{appa}.  The generalization concerns how to alter the original Hamiltonian towards a modified, control Hamiltonian that relies on the same chaotic (heteroclinic) trajectories as before, but is easier to implement numerically and, more importantly, experimentally.  The resulting quantum targeting technique is also fully coherent (unitary).  The original system's extreme sensitivity to initial conditions is still relied upon, but counteracting the quantum spreading (scrambling) by stabilizing the local dynamics is accomplished through the Hamiltonian modifications.  In this alternate approach, the Hamiltonian has locally stable dynamics, and yet it is still on an optimum track in order to arrive in an exponentially fast way and with relatively high precision at a predetermined target state and given time.  Here, the kicked rotor example in the main text is less precise than the earlier method, but adding an additional term in the Hamiltonian as described in Appendix~\ref{appb} would close that gap.  As before there are no free optimization parameters since everything is determined by the heteroclinic trajectory and the associated control Hamiltonian.  The general method is flexible in the sense that any trajectory of the original chaotic system can be selected and followed. In the specific case of the kicked rotor, increasing the precision only comes at the cost of increasing the complexity of the control Hamiltonian. This can be traced to the periodic nature of the rotor.  Ideally, the control kicks would have the appropriate sawtooth form in configuration space, and the increase in complexity comes from adding in additional harmonics in a Fourier series to better approximate the sawtooth shape.  In a dynamical system in which a time-dependent quadratic control Hamiltonian is found, this issue does not arise.  In fact, for the Bose-Hubbard system modeling ultracold bosonic atoms in an optical lattice such a quadratic Hamiltonian form appears~\cite{Beringer24} and this precision issue does not even appear.

As previously mentioned, the quantum kicked rotor has been realized in a broad range of ultracold atomic and other experiments~\cite{Moore95, Ammann98, Oberthaler99, Fischer99, Ryu06, Chabe08, Manai15, Bitter16, Hainaut18, Sajjad22, Cao22}, and for such experiments the only additional essential requirement for an experimental realization of optimal coherent chaotic quantum targeting is the ability to control magnitudes and phases of each kick. 

Finally, this approach may provide a conceptual platform for further applications to various systems in different branches of quantum control and quantum simulation, since it is devised in a rather general form.  In particular, an application of the ideas presented here to coherent states in quantum chaotic bosonic many-body systems exhibiting a semiclassical limit~\cite{Richter22} is immediate~\cite{Beringer24}. For example, choosing the target state identical to the initial state leads to a stabilized periodic many-body quantum dynamics (with any desired period or $(p,q)$ centroid).  This can be thought of as creating a nearly perfect quantum scarred state on a periodic orbit~\cite{Heller84}. 

Generalizing the concepts outlined here towards relevant many-body quantum control is left for exciting future work. 

\acknowledgments

We acknowledge support by the Deutsche Forschungsgemeinschaft (DFG, German Research Foundation), project Ri681/15-1, within the Reinhart-Koselleck Programme and Vielberth Foundation for financial support.  

\appendix
\section{Construction of $\widehat U_{M^{-1}}$}
\label{appa}

Part of the content of the semiclassical propagation method called linearized wave packet dynamics~\cite{Heller75, Heller91} is an accounting for wave packet spreading.  In other words, within that method there is a construction of a unitary operator in a configuration space representation that corresponds to a linear canonical transformation defined by a stability matrix, ${\bf M}$,
\begin{equation}
{\bf M} = \left( \begin{array}{cc} 
{\bf m}_{11} & {\bf m}_{21} \\ 
{\bf m}_{12} & {\bf m}_{22} \end{array} \right)  
\label{deltasm1}
\end{equation}
written in a block $2 \times 2$ form.   The ordering of components follows as in Eq.~\eqref{deltas2} (all momenta above the positions in the displacement vector).  

The generating function for the transformation, $S_\gamma({\bf q},{\bf q}^\prime)$, is at most a quadratic function of its arguments.  A constant term would represent a global phase and is irrelevant, and linear terms represent shift operators.  The important information therefore lies in the second derivatives, which satisfy the equation~\cite{Pal16}
\begin{align}
&\left(\begin{array}{cc}  \frac{\partial^2 S_\gamma }{\partial {\bf q} \partial {\bf q}} & \frac{\partial^2 S_\gamma }{\partial {\bf q}^\prime \partial {\bf q}} \\ \\ \frac{\partial^2 S_\gamma }{\partial {\bf q} \partial {\bf q}^\prime} & \frac{\partial^2 S_\gamma }{\partial {\bf q}^\prime \partial {\bf q}^\prime}\end{array}\right) =  \\
&\qquad \left( \begin{array}{cc} \bf{m_{11}}\cdot  \bf{m_{21}}^{-1} & \quad \bf{m_{12}} - \bf{m_{11}} \cdot  \bf{m_{21}}^{-1} \cdot  \bf{m_{22}} \\ 
- \bf{m_{21}}^{-1} & \bf{m_{21}}^{-1} \cdot \bf{m_{22}} \end{array} \right) 
\nonumber
\, .
\end{align}
For the optimal coherent targeting of Ref.~\cite{Tomsovic23}, it is necessary to use the inverse stability matrix to unwind the spreading of the dynamics.  Using block $2 \times 2$ identities for inverses~\cite{Lu02}, it can be shown that the generating function second derivatives must satisfy
\begin{align}
&\left(\begin{array}{cc}  \frac{\partial^2 S_\gamma }{\partial {\bf q} \partial {\bf q}} & \frac{\partial^2 S_\gamma }{\partial {\bf q}^\prime \partial {\bf q}} \\ \\ \frac{\partial^2 S_\gamma }{\partial {\bf q} \partial {\bf q}^\prime} & \frac{\partial^2 S_\gamma }{\partial {\bf q}^\prime \partial {\bf q}^\prime}\end{array}\right) = \\
&\qquad \left( \begin{array}{cc} -\bf{m_{21}^{-1}}\cdot  \bf{m_{22}} & \bf{m_{21}}^{-1}   \\ 
 \bf{m_{11}} \cdot  \bf{m_{21}}^{-1} \cdot  \bf{m_{22}} -\bf{m_{12}} \quad & -\bf{m_{11}} \cdot \bf{m_{21}^{-1}} \end{array} \right) \, .
 \nonumber 
\end{align}
At the moment in time, $t$, that the spreading is countered, the transformation must be centered locally about the heteroclinic trajectory's location ${\bf (q_t,p_t)}$, i.e.~using the function's arguments to be ${\bf q-q_t}$ and ${\bf q'-q_t}$.  Hence, the generating function is nearly, but not quite
\begin{widetext}
\begin{equation}
S_\gamma({\bf q},{\bf q}^\prime) =   \left( \begin{array}{cc} 
{\bf q-q_t} , & {\bf q'-q_t} \end{array} \right)  \left( \begin{array}{cc} -\bf{m_{21}^{-1}}\cdot  \bf{m_{22}} & \bf{m_{21}}^{-1}   \\ 
 \bf{m_{11}} \cdot  \bf{m_{21}}^{-1} \cdot  \bf{m_{22}} -\bf{m_{12}} \quad & -\bf{m_{11}} \cdot \bf{m_{21}^{-1}} \end{array} \right)  \left( \begin{array}{c} 
{\bf q-q_t} \\ 
{\bf q'-q_t} \end{array} \right)  \, .
\end{equation}
\end{widetext}
It turns out that the constructed unitary transformation
\begin{equation}
\bra{\bf q}\widehat U_{M^{-1}}\ket{\bf q^\prime} = \frac{{\rm Det}\left|\bf{m_{21}} \right|^{1/2} }{\left(2\pi i \hbar\right)^{D/2}}\exp\left[ \frac{i}{\hbar} S_\gamma({\bf q},{\bf q}^\prime)\right] \ ,
\end{equation}
 where $D$ is the number of degrees of freedom, does not center the momentum coordinate properly.  However, this is straightforwardly accomplished by multiplying the transformation above by $\exp\left[i {\bf p_t\left(q-q'\right)}/\hbar\right]$.  


\section{Constructing efficient control Hamiltonians for the kicked rotor}
\label{appb}

To begin with, assume that one would prefer to just add a small kick at the half time intervals between kicks in the original kicked rotor.  Such a time-dependent Hamiltonian takes the form
\begin{equation}
\label{krgc2}
\begin{split}
H_\gamma(p,q;t) = \frac{p^2}{2} + &\sum_{n=-\infty}^\infty  \bigg[ \delta(t-n) V_\gamma(q;n) + \\
&   \left. \delta\left(t-n-\frac{1}{2}\right) {\cal V}_\gamma\left(q;{n+\frac{1}{2}}\right) \right]
 \end{split}
\end{equation}
and the goal is to determine the two kicking potentials, $V_\gamma(q;n)$ and ${\cal V}_\gamma\left(q;{n+\frac{1}{2}}\right)$.  The idea is to use the constraints that Eq.~\eqref{eq:two}
holds for the initial condition $\left[p_\gamma(0),q_\gamma(0)\right]$ and that the one time step stability matrix is stable, which in this case ideally means it is a rotation matrix.  Hence $m_{11}^2+m_{21}^2=1$, $m_{11}=m_{22}$, and $m_{12}=-m_{21}$. 

Using the appropriate ordering of the matrices
\begin{equation}
{\bf M}_n={\bf M}_f{\bf M}_{K_{n+\frac{1}{2}}}{\bf M}_f{\bf M}_{K_n} 
\end{equation}
with
\begin{align}
{\bf M}_f &= \left( \begin{array}{cc} 
1 & 0 \\ 
\frac{1}{2} & 1 \end{array} \right) \nonumber \\
{\bf M}_{K_n} &= \left( \begin{array}{cc} 1 & - V_\gamma^{\prime\prime}(q;n) \\
 0 & 1 \end{array} \right) \nonumber \\
 {\bf M}_{K_{n+\frac{1}{2}}} &= \left( \begin{array}{cc} 1 & -{\cal V}_\gamma^{\prime\prime}\left(q;{n+\frac{1}{2}}\right)  \\
 0 & 1 \end{array} \right)
\label{stabm}
\end{align}
the one full time step stability matrix elements turn out to be given by
\begin{align}
m_{11} & = 1 - \frac{{\cal V}_\gamma^{\prime\prime}\left(q;{n+\frac{1}{2}}\right)}{2}\nonumber \\
m_{21} & = 1 - \frac{{\cal V}_\gamma^{\prime\prime}\left(q;{n+\frac{1}{2}}\right)}{4} \nonumber \\
m_{12} & = - {\cal V}_\gamma^{\prime\prime}\left(q;{n+\frac{1}{2}}\right) - V_\gamma^{\prime\prime}(q;n)  \nonumber \\ & \qquad\qquad\qquad\qquad\qquad + \frac{{\cal V}_\gamma^{\prime\prime}\left(q;{n+\frac{1}{2}}\right) V_\gamma^{\prime\prime}(q;n)}{2} \nonumber \\
m_{22} & = 1 - V_\gamma^{\prime\prime}(q;n) - \frac{{\cal V}_\gamma^{\prime\prime}\left(q;{n+\frac{1}{2}}\right)}{2} \nonumber \\ & \qquad\qquad\qquad\qquad\qquad  + \frac{{\cal V}_\gamma^{\prime\prime}\left(q;{n+\frac{1}{2}}\right) V_\gamma^{\prime\prime}(q;n)}{4} \, . \nonumber 
\end{align}

There are two solutions, say ${\cal A}$ and ${\cal B}$ respectively, satisfying the above mentioned  rotation matrix constraints, 
\begin{align}
V^{\prime\prime}\left[q_\gamma(n);n\right] & = 0 \nonumber
\\
{\cal V}_\gamma^{\prime\prime}\left[q_\gamma\left(n+\frac{1}{2}\right);{n+\frac{1}{2}}\right] & = \frac{4}{5}
\label{eq:A}
\end{align}
and
\begin{align}
V^{\prime\prime}\left[q_\gamma(n);n\right] & = 4 \nonumber \\
{\cal V}_\gamma^{\prime\prime}\left[q_\gamma\left(n+\frac{1}{2}\right);{n+\frac{1}{2}}\right] & = 4 \, .
\label{eq:B}
\end{align}
Note that the derivative relations hold for the control trajectories, but it would be helpful if the relations held in as large a local region as possible.  It is therefore quite useful to consider Taylor series expansions about the points $q_\gamma(n)$ and $q_\gamma\left(n+\frac{1}{2}\right)$ with respect to the constraints.  

It is also essential to maintain the periodicity of the potentials since its absence would lead to a non-rotor system. It is useful to imagine the potentials to be expanded in discrete sine and cosine Fourier series and given the Taylor series argument above, the arguments of the sine and cosine functions are ideally $2\pi n\left[q-q_\gamma(n)\right]$ and $2\pi n\left[q-q_\gamma\left(n+\frac{1}{2}\right)\right]$ for the integer and half integer kicks respectively.  The integer $n$ is the usual Fourier series index.

The other part of the constraint, preserving Eq.~\eqref{eq:two}, leads to the requirements that
\begin{align}
V_\gamma^{\prime}\left[q_\gamma(n);n\right] & = \frac{K}{2\pi}\sin\left[2\pi q_\gamma(n)\right] \nonumber \\
{\cal V}_\gamma^{\prime}\left[q_\gamma\left(n+\frac{1}{2}\right);{n+\frac{1}{2}}\right] & =  0
\end{align}
and hence, the equations of motion and stability requirements give values for the first and second derivatives of the potential in the control Hamiltonian on the trajectory.  Furthermore, a direct consequence of using the difference arguments in the Fourier series means that sine terms only contribute to the mapping equations whereas the cosine terms contribute only to the stability equations.

The trajectories in the immediate neighborhood of the control trajectory do not necessarily satisfy the constraints above exactly.  The most critical deviations to be controlled are in the stability relations.  In essence, if a significant neighborhood of the control trajectory remains stable (rotational), then small non-vanishing terms in the mapping do not harm the control procedure.  This idea is applied ahead to calculate additional terms in the potentials improving the accuracy.

\begin{figure}
\includegraphics[width=0.9\columnwidth]{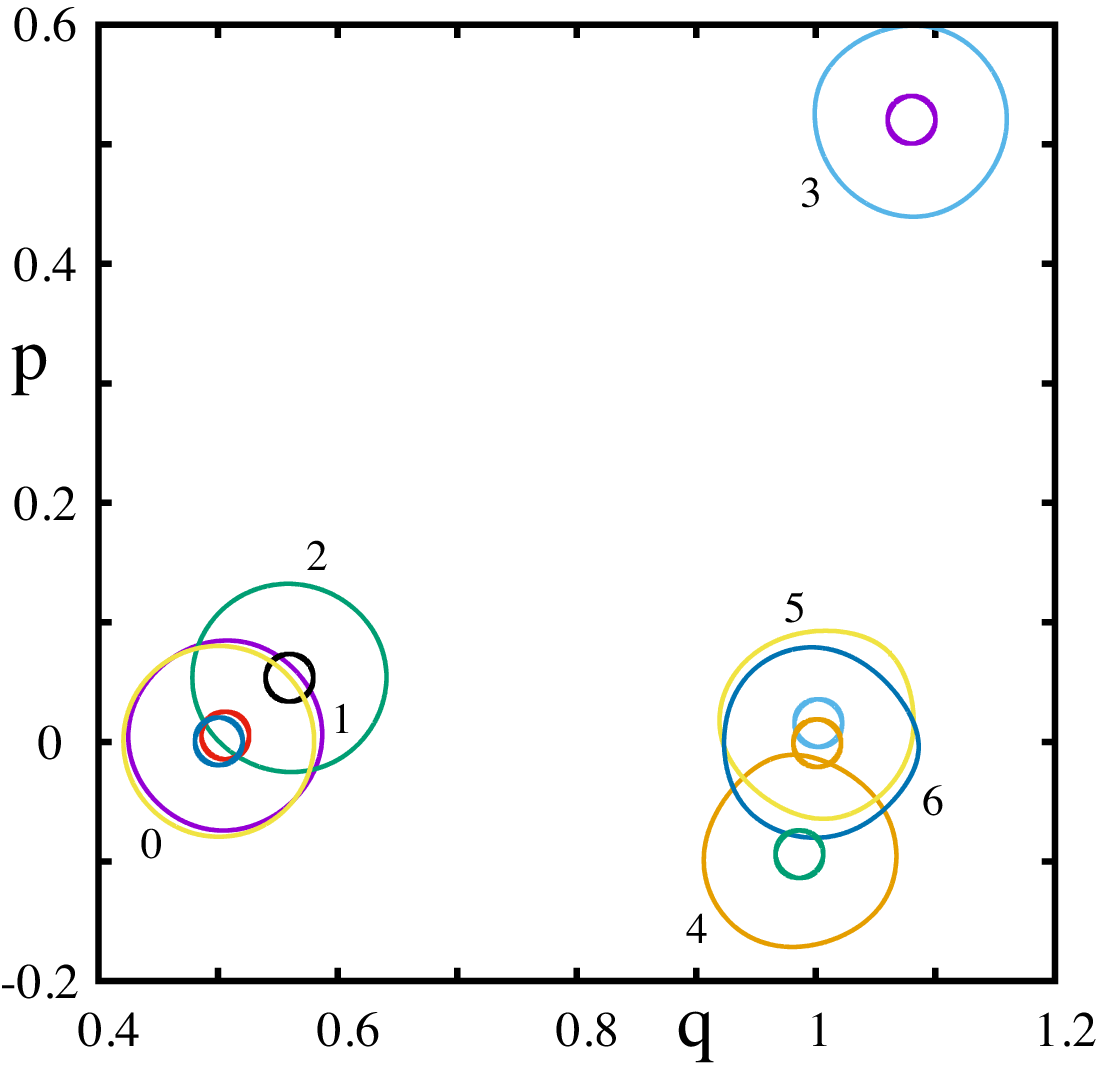}
\caption{Evolution of the Wigner transform densities of the propagated initial state for the Hamiltonian of Eq.~\eqref{krgc2} using the modification of Eq.~\eqref{mod2} in Eq.~\eqref{potential}.  The numbers label the time for each propagated density.  The outer circle is the $2\sigma$ contour, which encloses an area $h$, i.e.~that area occupied by a single quantum state.  The $\sigma/2$ contour is also drawn.  \label{twa2}}
\end{figure}

\subsection{Solution ${\cal A}$}

Solution ${\cal A}$, obeying Eq.~(\ref{eq:A}),
leads to a simpler control Hamiltonian with a weaker perturbation at the half time steps.  It is thus preferable to solution ${\cal B}$, satisfying Eq.~(\ref{eq:B}), and it is considered first.  The simplest potentials, i.e. involving only the first terms in the Fourier series, consistent to leading order in $q-q_\gamma(n)$ and $q-q_\gamma\left(n+\frac{1}{2}\right)$ with the constraints just discussed are
\begin{align}
V_\gamma\left(q;n\right) & = \frac{K}{4\pi^2}\sin\left[2\pi q_\gamma(n)\right]\sin\left(2\pi\left[q-q_\gamma(n)\right]\right) \nonumber \\
{\cal V}_\gamma\left(q;{n+\frac{1}{2}}\right) & =  -\frac{1}{5\pi^2}\cos\left(2\pi \left[q-q_\gamma\left(n+\frac{1}{2}\right)\right]\right)
\label{potential}
\end{align}
and this is the example control Hamiltonian used in the main text.  It works fairly well, though not as well as the original control relying on $\widehat U_{M^{-1}}$ of Ref.~\cite{Tomsovic23}; see Fig.~\ref{overlap2}.

\begin{figure}
\includegraphics[width=0.9\columnwidth]{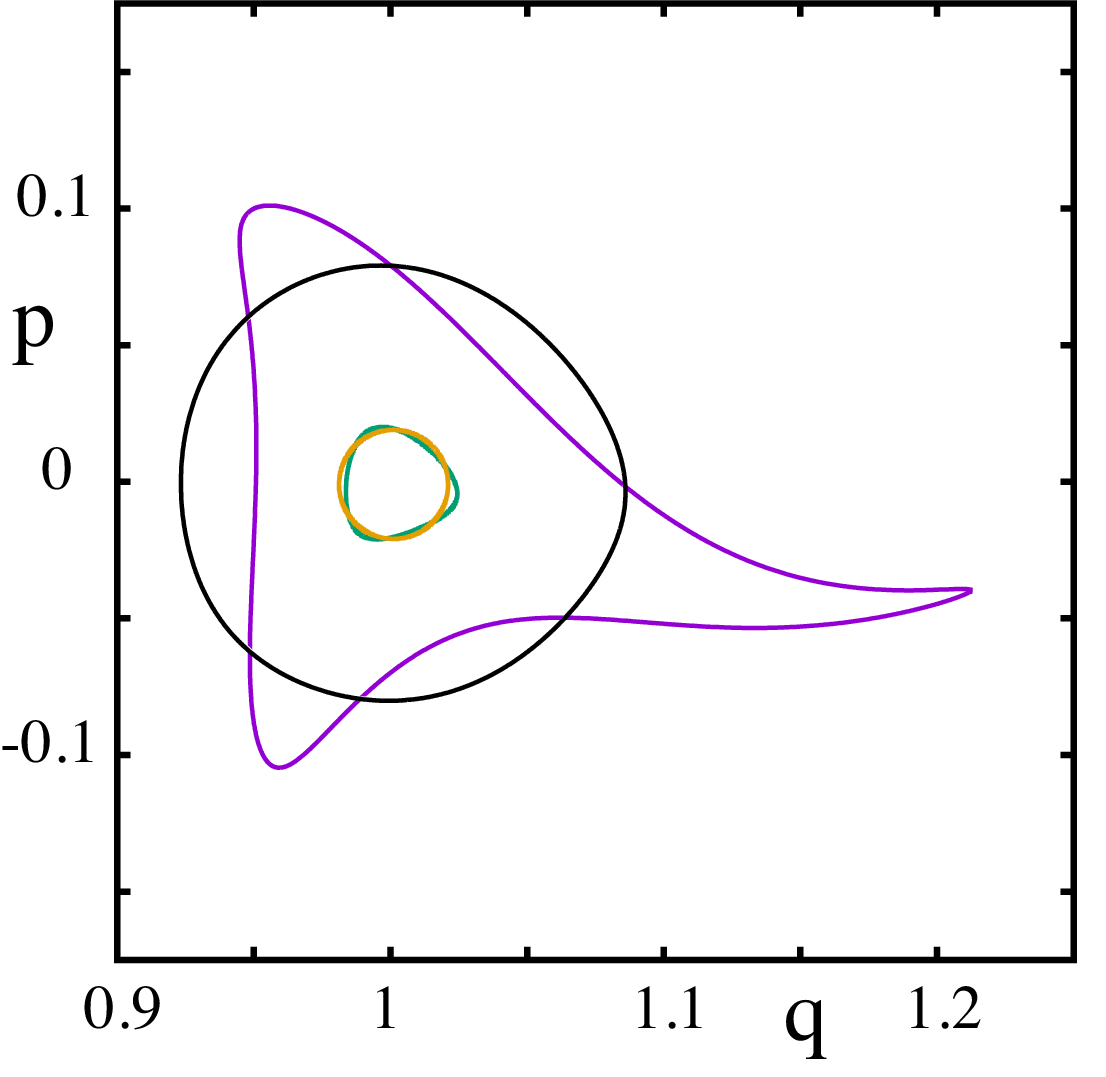}
\caption{Comparison of the final propagated Wigner transform densities for the Hamiltonians of Eq.~\eqref{krgc} and Eq.~\eqref{krgc2} using the modification of Eq.~\eqref{mod2} in Eq.~\eqref{potential}.  The outer circles are the $2\sigma$ contours, which show a much greater improvement than for the $\sigma/2$ contours.  \label{twacomp}}
\end{figure}

However, notice that the $V_\gamma\left(q;n\right)$ potential generates leading order corrections to the stability matrix elements $m_{12}$ and $m_{22}$ proportional to $q-q_\gamma(n)$ and they are multiplied by the large factor $K$.  This is responsible for the increasing distortion with time seen in Fig.~\ref{twa} of the Wigner densities.  They are not remaining circularly symmetric with increasing time and, in fact, the linear deviations from zero of neighboring trajectories in the second derivative of $V_\gamma\left(q;n\right)$ are the dominant reason.  By considering the higher order term of double the frequency, it is possible to cancel this linear term in the second derivative of the potential.  It is straightforward to show that the replacement of 
\begin{align}
&\sin\left(2\pi\left[q-q_\gamma(n)\right]\right)  \ {\rm with} \nonumber \\
&\qquad\quad \frac{4}{3}\left[\sin\left(2\pi\left[q-q_\gamma(n)\right]\right) - \frac{1}{8}\sin\left(4\pi\left[q-q_\gamma(n)\right]\right)\right]
\label{mod2}
\end{align}
in $V_\gamma\left(q;n\right)$ eliminates this linear term yet leaves Eq.~\eqref{modifiedM} unchanged.  With this frequency doubled term in the
potential, the analogous version of Fig.~\ref{twa} becomes Fig.~\ref{twa2}; see also the direct comparison and improvement of the final densities in Fig.~\ref{twacomp}.  The densities remain much closer to their original circular forms.  This indicates that using this improved control Hamiltonian would achieve a greater accuracy than the one used in the main text for a given value of $\hbar$; see Fig.~\ref{overlap2} for the unimproved case.  Continuing with the logic presented here, it would be possible to create an infinite sequence of ever improving control Hamiltonians canceling higher and higher order corrections, but the increasing complexity of the Hamiltonians is unlikely to merit the minor additional accuracy in the control process.

\begin{figure}[ht]
\includegraphics[width=0.9\columnwidth]{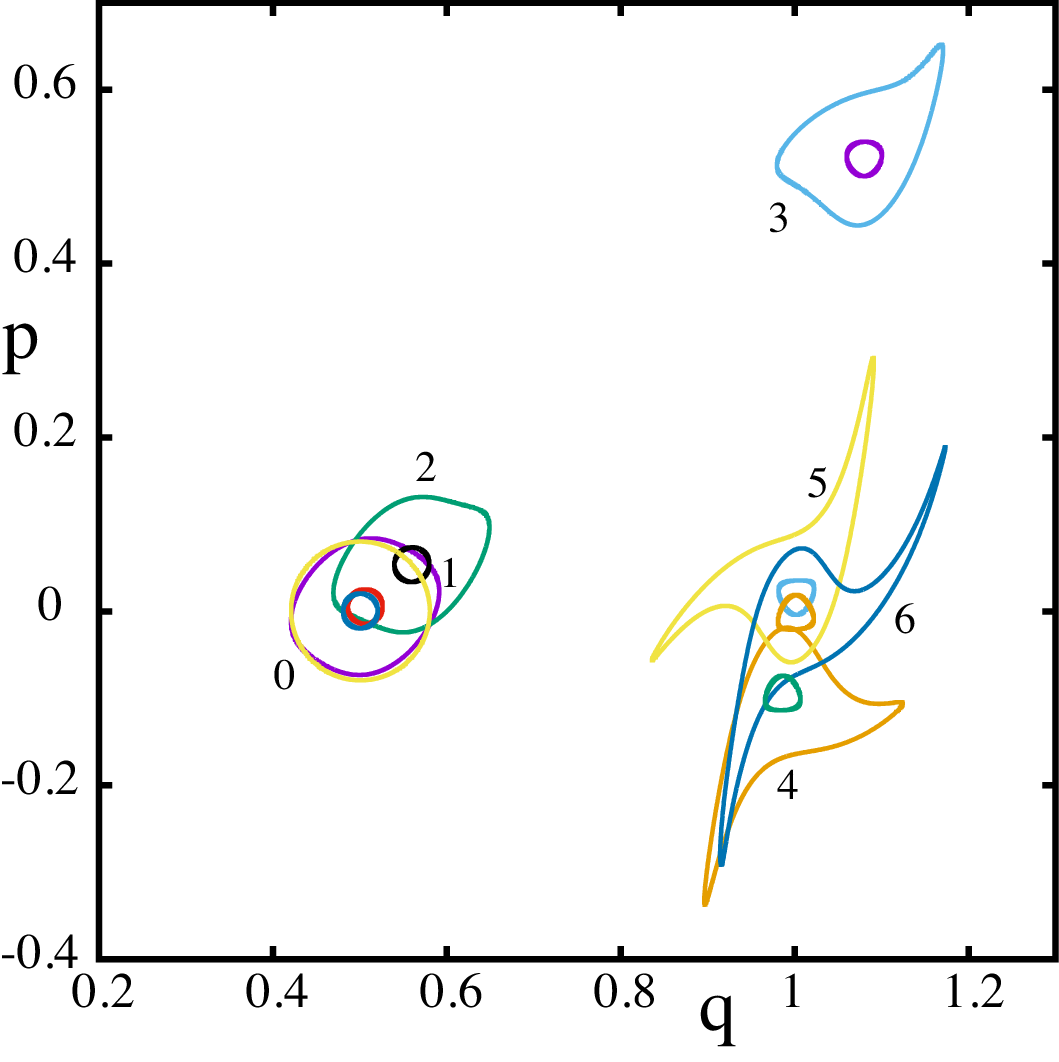}
\caption{Evolution of the Wigner transform densities of the identical propagated initial state, Fig.~\ref{fig2}, under the dynamics of Eq.~\eqref{krgc2} with the potentials in Eqs.~(\ref{eqbv1},\ref{krv1p}).  The outer circle is the $2\sigma$ contour, which encloses an area $h$, i.e.~that area occupied by a single quantum state.  The $\sigma/2$ contour is also drawn.  As previously, $K=8$, and $N=200$.  \label{twa3}}
\end{figure}

\subsection{Solution ${\cal B}$}

This solution is more complicated because the second derivative for the integer kick potential does not lead to a vanishing contribution to the stability, and yet is responsible for the non-vanishing term in the mapping.  Thus, it must incorporate both a sine and cosine term even in the simplest case.  The term can be written in a slightly more compact way by combining the terms into a single phase shifted cosine term.  This gives
\begin{align}
V_\gamma(q;n) & = -\frac{1}{\pi^2}\cos\left(2\pi \left[q-q_\gamma(n)\right] \right) + \nonumber \\ & \qquad \qquad \frac{K\sin\left[2\pi q_\gamma(n)\right]}{4\pi^2} \sin\left(2\pi \left[q-q_\gamma(n)\right] \right) \nonumber \\
& = -\frac{K(n)}{4\pi^2}\cos\bigg(2\pi \left[q-q_\gamma(n)\right]+\phi(n)\bigg) \nonumber
\end{align}
with 
\begin{align}
K(n) & = K \sqrt{\frac{16}{K^2}+\sin^2\left[2\pi q_\gamma(n)\right]} \nonumber\\
\phi(n) & = \sin^{-1}\left[\frac{\sin\left[2\pi q_\gamma(n)\right]}{\sqrt{\frac{16}{K^2}+\sin^2\left[2\pi q_\gamma(n)\right]}}\right]
\label{eqbv1}
\end{align}
and
\begin{equation}
\label{krv1p}
{\cal V}_\gamma\left(q;{n+\frac{1}{2}}\right) = - \frac{1}{\pi^2}\cos\left(2\pi \left[q-q_\gamma\left(n+\frac{1}{2}\right)\right]\right)\, .
\end{equation}
For the control heteroclinic trajectory this leads to the stability matrix
\begin{equation}
{\bf M}_n(\gamma) = \left( \begin{array}{cc} 
-1 & 0  \\ \\
0  & -1  \end{array} \right) = \left( \begin{array}{cc} 
1 & -4  \\ \\
\frac{1}{2} & -1  \end{array} \right)\left( \begin{array}{cc} 
1 & -4  \\ \\
\frac{1}{2} & -1  \end{array} \right) \, .
\label{modifiedM2}
\end{equation}
In this case, the stability matrix is that of a rotation about $\pi$ radians.

This solution is less desirable, not only because it is more complicated and requires a five times stronger perturbation at half integer times steps, but also because it leads to worse distortion of the Wigner transform densities locally for a given $\hbar$.   Nevertheless, trajectories in a small neighborhood around the control trajectory remain relatively stable for some time.  This is illustrated in Fig.~\ref{twa3}, which can be compared to Fig.~\ref{twa} and Fig.~\ref{twa2}.

This protocol is somewhat worse at maintaining locally stable dynamics than the example treated in the main text.  However, just as for the case of solution ${\cal A}$, making the substitution of Eq.~\eqref{mod2} appropriately for $V_\gamma(q;n)$ in Eq.~\eqref{eqbv1} decreases the deformation away from circular densities.  It is not shown here due to its similar improvement as in the case of solution ${\cal A}$.


\bibliography{classicalchaos,furtherones,general_ref,molecular,quantumchaos,rmtmodify,manybody}

\end{document}